\documentclass[aps,pra, twocolumn,showpacs]{revtex4-1}

\usepackage{graphicx}
\usepackage{amsmath, calc}
\usepackage{amsfonts}
\usepackage{amssymb}
\usepackage{bm}
\usepackage{bbm}
\usepackage{color}
\usepackage{array}
\usepackage{times}
\usepackage{units}
\usepackage{comment}

\usepackage[pdftex, colorlinks=true, linkcolor=blue, citecolor=blue, urlcolor=blue]{hyperref}

\newcommand{\bra}[1]{\langle #1|}
\newcommand{\ket}[1]{|#1\rangle}

\newcommand{\expect}[1]{\langle #1 \rangle}
\DeclareMathOperator{\sgn}{sgn}

\begin{document}

\title{Topological collective plasmons in bipartite chains of metallic nanoparticles}

\author{Charles A. Downing}
\affiliation{Universit\'{e} de Strasbourg, CNRS, Institut de Physique et Chimie des Mat\'eriaux de Strasbourg, UMR 7504, F-67000 Strasbourg, France}

\author{Guillaume Weick}
\email{guillaume.weick@ipcms.unistra.fr} 
\affiliation{Universit\'{e} de Strasbourg, CNRS, Institut de Physique et Chimie des Mat\'eriaux de Strasbourg, UMR 7504, F-67000 Strasbourg, France}


\begin{abstract}
We study a bipartite linear chain constituted by spherical metallic nanoparticles, where each nanoparticle supports a localized surface plasmon. The near-field dipolar interaction between the localized surface plasmons gives rise to collective plasmons, which are extended over the whole nanoparticle array. We derive analytically the spectrum and the eigenstates of the collective plasmonic excitations. At the edge of the Brillouin zone, the spectrum is of a 
pseudorelativistic nature similar to that present in the electronic band structure of polyacetylene. We find the effective Dirac Hamiltonian for the collective plasmons and show that the corresponding spinor eigenstates represent one-dimensional Dirac-like massive bosonic excitations. Therefore, the plasmonic lattice exhibits similar effects to those found for electrons in one-dimensional Dirac materials, such as the ability for transmission with highly suppressed backscattering due to Klein tunneling. We also show that the system is governed by a nontrivial Zak phase, which predicts the manifestation of edge states in the chain. When two dimerized chains with different topological phases are connected, we find the appearance of the bosonic version of a Jackiw-Rebbi midgap state. We further investigate the radiative and nonradiative lifetimes of the collective plasmonic excitations and comment on the challenges for experimental realization of the topological effects found theoretically.

\end{abstract}

\maketitle

\section{Introduction}

Since Veselago \cite{Veselago1968} first pondered creating artificial materials to induce unorthodox physical properties in 1968, metamaterials have been at the forefront of interdisciplinary research to develop novel applications. After the turn of the new millennium, groundbreaking progresses on invisibility and cloaking \cite{Schurig2006}, slow-light \cite{Tsakmakidis2007}, and superlensing effects \cite{Zhang2008} have been reported. This scientific paradigm has spawned the subfield of nanostructure-based plasmonic metamaterials, which exploit the properties of the surface plasmon, a collective oscillation of valence electronic charges \cite{Bertsch1994, Kreibig1995, Maier2007}. 

The ability for plasmonic materials to focus light at subwavelength scales and to generate a large optical density of states are two significant physical properties which may be utilized in future applications \cite{Barnes2003}. There have already been numerous proposed implementations of plasmonic metamaterials, including spasers \cite{Bergman2003}, optical transmission modulators \cite{Pacifici2007}, bio-sensors \cite{Anker2008}, switches \cite{Pala2008} and optical storage devices \cite{Zijlstra2015}. There is continued and active interest in schemes enabling long-range propagation in plasmonic waveguides \cite{Stockman2004, Oulton2008} since they are the building blocks of plasmonic circuitry \cite{Barnes2003}, where it is hoped that a similar large bandwidth of information as in conventional photonics can be achieved, but without the limitation of diffraction at submicron cross sections.

Linear, one-dimensional (1D) chains of regularly-spaced metallic nanoparticles, which support localized surface plasmons (LSPs) confined to each nanoparticle, have been proposed almost two decades ago to act as plasmonic waveguides \cite{Quinten1998}. When irradiated with light, the LSP excitations in the nanoparticles forming the array couple to each other through the optical near-field dipole-dipole interaction, which induces collective plasmons \cite{Meinzer2014} that extend over the whole chain and which guide electromagnetic radiation with strong lateral confinement. Since the original proposal by Quinten \textit{et al.}\ \cite{Quinten1998}, plasmon propagation in nanoparticle chains has been extensively investigated both theoretically \cite{Brongersma2000, Maier2003b, Weber2004, Park2004, Citrin2004, Markel2007, Lee2012, Pino2014, Brandstetter2016} and experimentally 
\cite{Krenn1999, Maier2002, Maier2003a, Solis2012, Apuzzo2013, Barrow2014}. 

Recently, the potentially nontrivial topological properties of collective excitations in zigzag \cite{poddu14_ACSPhoton} and `diatomic' \cite{Ling2015} chains of nanoparticles 
were uncovered. 
In particular, it has been shown in Ref.\ \cite{Ling2015}  by means of a classical quasistatic calculation based on 
the macroscopic Maxwell equations that  
the collective plasmonic oscillations in the linear array become decidedly topologically nontrivial when the chain becomes dimerized. 
Other 1D or quasi-1D artificial systems such as photonic lattices \cite{Kanshu2012, Schomerus2013}, dielectric resonators \cite{Poli2015, Slobo2015}, silicon waveguides \cite{Blanco2016}, polariton cavities \cite{Solny2016, Solny2016b}, and metallic nanowires \cite{liu16_preprint} were also shown to present interesting nontrivial topological features, paving the way to a new exciting field of research coined `topological photonics' \cite{Lu2014}.

In this work, we present a fully quantum-mechanical treatment of collective plasmons in a linear bipartite chain of metallic nanoparticles, which provides a clear analogy with the nontrivial electronic properties of fermionic topological insulators \cite{hasan10_RMP}. Such an analogy remains only elusive within a classical treatment \cite{Ling2015}. Moreover, a quantum treatment becomes essential when the size of the nanoparticles composing the chain (below ca.\ $\unit[20]{nm}$ in diameter) is such that quantum-size effects appear in their optical response \cite{Bertsch1994, Kreibig1995, Tame2013}. 
The formed bipartite lattice, given by a regular array with a unit cell consisting of two identical nanoparticles, 
presents a small gap at the edge of the first Brillouin zone in the two-band collective plasmonic bandstructure. 
We find an effective Hamiltonian for the collective plasmons, which at the edge of the Brillouin zone corresponds to a 1D Dirac equation for massive bosonic quasiparticles. These Bogoliubov excitations are accompanied with a nontrivial vacuum state, which is different from the vacuum of the original plasmonic excitations. The Dirac-like nature of the quasiparticles associated with our system allows us to propose a setup to observe the plasmonic analogue of Klein tunneling \cite{Klein1929}, which is celebrated for enabling transmission with highly suppressed backscattering, as is the case in two-dimensional graphene sheets \cite{katsn06_NatPhys}.

We show that depending on the strength of the dimerization, the system may be in either a topologically trivial or nontrivial phase, which is codified by a topological invariant \cite{Qi2008}. This invariant governs the existence of special, topologically-protected modes at the ends of the chain and identifies the bipartite nanoparticle chain as a `bosonic topological insulator' for plasmonic excitations. We unveil several types of edge states, which appear at the midgap frequency of the gapped plasmonic bandstructure. We further investigate the topologically-protected midgap state formed between two dimerized chains, which is reminiscent of the Jackiw-Rebbi state \cite{Jackiw1976}. 

A major challenge for the effective use of plasmonic circuitry is how to effectively combat losses necessarily present in a system with a high scattering rate of electrons, radiation losses into the far-field, and dissipation due to quantum-size effects \cite{Khurgin2015}. Here we estimate, with analytical expressions for both radiation and Landau \cite{Kawabata1966, Bertsch1994, Kreibig1995} damping decay rates, the plasmonic losses in the bipartite chain. In particular, the collective plasmonic bandstructure is comprised, for a given polarization, of one band where the dipole moments on each nanoparticle within a dimer are in-phase, while on the other band these dipoles are out-of-phase. For the in-phase band, only those states within the light cone 
significantly couple to light and thus suffer from radiation losses. Such modes are thus referred to as `bright'. All the other plasmonic states weakly couple to light and are therefore immune from radiation damping. Hence these modes are called `dark'.
In contrast, all of the modes are subject to nonradiative losses, i.e., Ohmic and Landau damping. 
Our detailed analysis of the various damping mechanisms the collective plasmons suffer from 
enables us to comment on the experimental challenges to observe topological effects in nanoplasmonics.

The sequel of this paper is organized as follows: Section~\ref{sec2} presents our model of collective plasmonic excitations in a bipartite chain of metallic nanoparticles coupled to a bath of electron-hole pairs (responsible for the nonradiative Landau damping) as well as to a photonic bath (which leads to radiation damping of the collective excitations). In Sec.~\ref{sec3}, we diagonalize the purely plasmonic problem and discuss the properties of the associated plasmonic bandstructure. We detail in Sec.~\ref{sec4} the mapping of the plasmonic Hamiltonian to a 1D Dirac equation for massive bosonic quasiparticles, and consider the plasmonic version of Klein tunneling. In Sec.~\ref{sec5}, we study the topological properties of the bipartite chain, codified by a nontrivial Zak phase, and illustrate how various types of edge states and topologically-protected midgap states may form. An analysis of plasmonic losses due to both the size-dependent radiation and Landau damping is presented in Sec.~\ref{sec6}, which enables us to comment on the experimental realization of our proposal. Finally, we draw conclusions in Sec.~\ref{sec7}.
A few appendices complement the discussion presented in the main text.

\section{Model}
\label{sec2}

\begin{figure}[tb]
 \includegraphics[width=1.00\columnwidth]{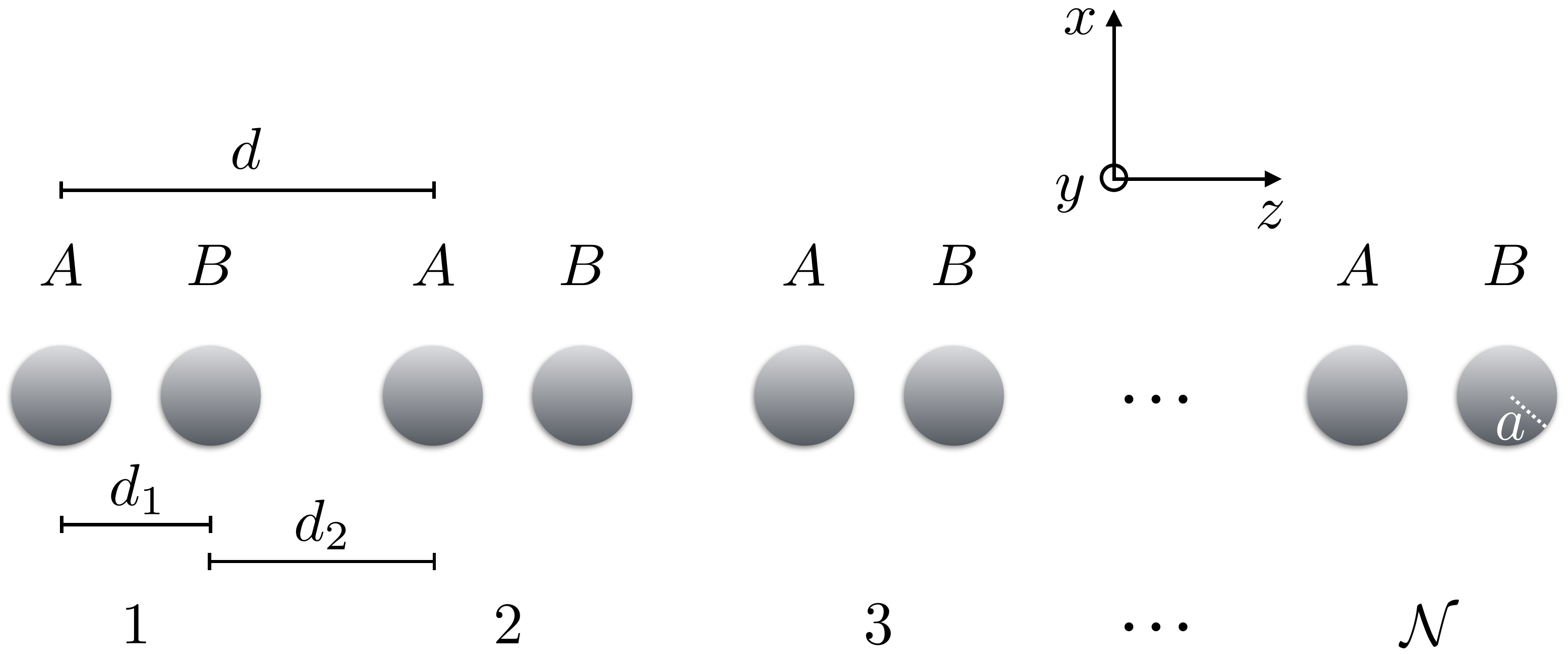}
 \caption{Sketch of a chain of $\mathcal{N}$ dimers, comprised of a pair of metallic nanoparticles denoted $A$ and $B$, each supporting localized surface plasmon resonances.}
 \label{fig1}
\end{figure}

We consider a linear chain composed of $\mathcal{N}$ dimers, which comprise the unit cell of two identical spherical metallic nanoparticles. The two 
corresponding inequivalent sublattices are denoted by $A$ and $B$, respectively. This geometry is sketched in Fig.\ \ref{fig1}, showing the nanoparticle radius $a$ and the interparticle separations $d_{1}$ and $d_{2}$ which build a bipartite chain with a period $d = d_1 + d_2$. Along the chain each nanoparticle supports an LSP resonance \cite{Bertsch1994, Kreibig1995, Maier2007}, a collective excitation of the electronic center of mass, which couples to the neighboring nanoparticles via the dipole-dipole interaction \cite{Brongersma2000}. This approximation is valid for interparticle distances $d_{1, 2}\gtrsim 3a$ \cite{Park2004}. In what follows, we denote by $H_{\mathrm{pl}}$ the plasmonic Hamiltonian describing the near-field-coupled LSPs along the bipartite chain.

Each LSP is coupled to vacuum photonic modes described by the Hamiltonian $H_\mathrm{ph}$ through the light-matter interaction $H_{\mathrm{pl}\textrm{-}\mathrm{ph}}$ (which includes retardation effects), leading to the radiative decay of the collective excitations. 
Moreover, the coupling described by the Hamiltonian $H_{\mathrm{pl}\textrm{-}\mathrm{eh}}$ of the LSPs to electron-hole pairs (with Hamiltonian $H_{\mathrm{eh}}$) in each of the nanoparticles leads to the size-dependent Landau damping of the collective modes \cite{Brandstetter2016}. 
The Hamiltonian of our system hence reads
\begin{equation}
\label{eq:H}
H=H_{\mathrm{pl}}+H_\mathrm{ph}+H_{\mathrm{pl}\textrm{-}\mathrm{ph}}+H_{\mathrm{eh}}+H_{\mathrm{pl}\textrm{-}\mathrm{eh}}.
\end{equation}

Since the nanoparticles we consider are small enough to be subject to quantum-size effects \cite{Bertsch1994, Kreibig1995, Kawabata1966, Tame2013}, and to elucidate the connection to fermionic quasiparticles in Dirac materials, we quantize the classical plasmonic Hamiltonian $H_{\mathrm{pl}}$ \cite{Weick2013, Weick2015, Sturges2015, Lamowski2016, Brandstetter2016} in the Coulomb gauge \cite{cohen, craig},  leading to 
\begin{align}
\label{eq01}
  H_{\mathrm{pl}} =&\; \hbar \omega_0 \sum_{n=1}^\mathcal{N} \sum_{\sigma = x, y, z} \left( {a_n^{\sigma}}^{\dagger} a_n^{\sigma} + {b_n^{\sigma}}^{\dagger}   b_n^{\sigma} \right) \nonumber\\
&+ \hbar \Omega_1  \sum_{n=1}^\mathcal{N} \sum_{\sigma = x, y, z} \eta_{\sigma} \left( {a_n^{\sigma}}^{\dagger}  + a_n^{\sigma} \right) \left(  {b_n^{\sigma}}^{\dagger}  + b_n^{\sigma} \right)  \nonumber \\
  &+ \hbar \Omega_2  \sum_{n=1}^{\mathcal{N}-1 } \sum_{\sigma = x, y, z} \eta_{\sigma} \left( {a_{n+1}^{\sigma}}^{\dagger}  + a_{n+1}^{\sigma} \right) \left( {b_n^{\sigma}}^{\dagger}  + b_n^{\sigma} \right), 
 \end{align}
where $n$ is an index which denotes the dimer number in the chain (see Fig.~\ref{fig1}), and the summation over $\sigma = x, y, z$ counts the two transverse ($x, y$) and the single longitudinal ($z$) polarizations of the collective plasmonic modes. The factor $\eta_x = \eta_y = 1$ ($\eta_z = -2$) accounts for the repulsive (attractive) interaction between transverse (longitudinal) nearest neighbor LSPs. We neglect interactions beyond those between nearest neighbors since such interactions only lead to a small quantitative change of the plasmonic dispersion (see Appendix \ref{appendA} for details).
The bosonic operators $a_n^{\sigma}$ and $b_n^{\sigma}$ (${a_n^{\sigma}}^{\dagger}$ and ${b_n^{\sigma}}^{\dagger}$) in Eq.~\eqref{eq01} annihilate (create) an LSP with polarization $\sigma$ in the $n$th dimer in nanoparticle $A$ and $B$, respectively. 

The Hamiltonian~\eqref{eq01} displays terms which are reminiscent of a 1D tight-binding model. Firstly, the role of the onsite energy is played by the LSP resonance frequency $\omega_0$ [here we refer to the first and second terms on the right-hand side of Eq.~\eqref{eq01}]. For the case of alkaline nanoparticles in vacuum, this resonance frequency corresponds to the Mie frequency $\omega_\mathrm{p}/3^{1/2}=(N_{\mathrm{e}} e^2 / m_{\mathrm{e}} a^3 )^{1/2}$, where $e$ is the charge of an electron with effective mass $m_{\mathrm{e}}$ and the number of electrons in each nanoparticle is $N_{\mathrm{e}}$ \cite{Kreibig1995}. 
The plasma frequency is $\omega_\mathrm{p}=(4\pi n_\mathrm{e}e^2/m_\mathrm{e})^{1/2}$, 
where $n_\mathrm{e}$ is the electronic density of the considered metal. 
Secondly, the analogue of the hopping terms in Eq.~\eqref{eq01}, namely those $\propto ( {a_n^{\sigma}}^{\dagger} b_n^{\sigma} + \mathrm{h.c.} )$ and $\propto ( {a_{n+1}^{\sigma}}^{\dagger} b_n^{\sigma} + \mathrm{h.c.} )$, are due to the nearest neighbor dipole-dipole interaction. The dipolar interaction gives rise to the coupling constants 
\begin{equation}
\label{eq:Omega}
\Omega_{1, 2} = \frac{\omega_0}{2} \left(\frac{a}{d_{1, 2}}\right)^3,
\end{equation}
which alternate along the chain. However, this is where the comparison to a standard tight-binding model ends, due to the appearance of nonresonant terms $\propto (a_n^{\sigma} b_n^{\sigma} + \mathrm{h.c.})$ and $\propto (a_{n+1}^{\sigma} b_n^{\sigma} + \mathrm{h.c.})$ which enter the Hamiltonian \eqref{eq01} and are also due to the dipolar interaction. Such nonrotating-wave terms cannot be neglected when calculating plasmon eigenstates and, hence, quantities dependent on the latter, such as damping rates (see Sec.\ \ref{sec6}) \cite{Brandstetter2016}.

Without the nonresonant terms mentioned above, the model encapsulated in the Hamiltonian \eqref{eq01} can be seen as the bosonic counterpart to the Su-Schrieffer-Heeger model \cite{Su1979, Su1980}, where electrons hopping in 1D 
lattices with staggered hopping amplitudes were considered. This simple, though physically rich model was originally developed in studies of conducting polymers such as polyacetylene and is known to exhibit rather exotic physics including fractionally charged excitations and edge states \cite{Jackiw1976}. How the presence of the bosonic nonresonant terms in the Hamiltonian \eqref{eq01} will affect the formation and properties of edge states is not immediately clear. This issue is addressed in the sequel of our paper. 

In Eq.\ \eqref{eq:H}, the vacuum modes of the electromagnetic field, to which the LSPs are coupled to, lead to the radiative damping of the plasmonic collective modes and are described by the photonic Hamiltonian 
\begin{equation}
\label{eq00343401}
 H_{\mathrm{ph}} = \sum_{\mathbf{k}, \hat{\lambda}_{\mathbf{k}}} \hbar \omega_{\mathbf{k}} {c_{\mathbf{k}}^{\hat{\lambda}_{\mathbf{k}}}}^{\dagger} c_{\mathbf{k}}^{\hat{\lambda}_{\mathbf{k}}},
\end{equation}
where $c_{\mathbf{k}}^{\hat{\lambda}_{\mathbf{k}}}$ (${c_{\mathbf{k}}^{\hat{\lambda}_{\mathbf{k}}}}^{\dagger}$)  annihilates (creates) a photon with momentum $\mathbf{k}$, transverse polarization $\hat{\lambda}_{\mathbf{k}}$, and dispersion $\omega_{\mathbf{k}} = c |\mathbf{k}|$, with $c$ the speed of light in vacuum. 
In the long-wavelength limit, the plasmon-photon coupling Hamiltonian entering Eq.\ \eqref{eq:H} has the expression
\begin{align}
\label{eq0003}
H_{\mathrm{pl}\textrm{-}\mathrm{ph}} =&\; 
\mathrm{i} \hbar \sum_{n=1}^{\mathcal{N}} \sum_{\sigma=x,y,z}\sum_{\mathbf{k}, \hat{\lambda}_{\mathbf{k}}} \sqrt{\frac{\pi (\omega_0 a)^3 }{ \mathcal{V} \omega_{\mathbf{k}} }} \hat{\sigma} \cdot \hat{\lambda}_{\mathbf{k}}  
\nonumber\\
&\times   \left[  \left( {a_n^{\sigma}}^{\dagger} - a_n^{\sigma} \right) \left( c_{\mathbf{k}}^{\hat{\lambda}_{\mathbf{k}}} \mathrm{e}^{\mathrm{i} \mathbf{k} \cdot \mathbf{d}_{n, A} } + {c_{\mathbf{k}}^{\hat{\lambda}_{\mathbf{k}}}}^{\dagger} \mathrm{e}^{-\mathrm{i} \mathbf{k} \cdot \mathbf{d}_{n, A} } \right)  
\right.\nonumber\\
& + \left.  \left( {b_n^{\sigma}}^{\dagger} - b_n^{\sigma} \right) \left( c_{\mathbf{k}}^{\hat{\lambda}_{\mathbf{k}}} \mathrm{e}^{\mathrm{i} \mathbf{k} \cdot \mathbf{d}_{n, B} } + {c_{\mathbf{k}}^{\hat{\lambda}_{\mathbf{k}}}}^{\dagger} \mathrm{e}^{-\mathrm{i} \mathbf{k} \cdot \mathbf{d}_{n, B} } \right) \right], 
\end{align}
where the position of the center of the nanoparticle belonging to sublattice $A$ ($B$) in the $n$th dimer is denoted by 
$\mathbf{d}_{n, A} = (n-1) d\, \hat{z}$ ($\mathbf{d}_{n, B} = \mathbf{d}_{n, A}+d_1\, \hat{z}$), and where $\mathcal{V}$ is the quantization volume of the 
photonic modes. We note that the coupling \eqref{eq0003}, together with the static, instantaneous interaction in Eq.\ 
\eqref{eq01}, results in a fully-retarded description of the dipole-dipole interaction between LSPs along the bipartite chain \cite{craig}.

The Hamiltonian describing the bath of electron-hole pairs in Eq.\ \eqref{eq:H} reads as
\begin{equation}
\label{eq0009}
 H_{\mathrm{eh}} = \sum_{n=1}^{\mathcal{N}} \sum_{s = A, B} \sum_{\mu} \varepsilon_{n s \mu} f_{n s \mu}^{\dagger} f_{n s \mu}^{\phantom{\dagger}},
\end{equation}
where $\mu$ labels the electron and hole states  with energy $\varepsilon_{n s \mu}$ in nanoparticle $(n,s)$ and where 
$f_{n s \mu} \left( f_{n s \mu}^{\dagger} \right)$ annihilates (creates) the one-body state $\ket{n s \mu}$. In the hard-wall approximation for the mean-field potential felt by the electrons \cite{Yannouleas1992, Weick2005}, the collective plasmon-electron-hole coupling leading to the Landau damping of the collective plasmons reads \cite{Brandstetter2015, Brandstetter2016}
\begin{align}
\label{eq0010}
 H_{\mathrm{pl}\textrm{-}\mathrm{eh}} =&\; 
 \Lambda \sum_{n=1}^\mathcal{N} \sum_{\sigma=x,y,z} \sum_{\mu \mu'}  \Big[  \left( a_n^{\sigma} + a_n^{\sigma \dagger} \right) \bra{n A \mu} \sigma \ket{n A \mu'}
 \nonumber\\
 &\times
  f_{n A \mu}^{\dagger} f_{n A \mu'}^{\phantom{\dagger}}
 \nonumber\\
  &+  \left( b_n^{\sigma} + b_n^{\sigma \dagger} \right) \bra{n B \mu} \sigma \ket{n B \mu'} f_{n B \mu}^{\dagger} f_{n B \mu'}^{\phantom{\dagger}} \Big],
\end{align}
where $\Lambda = \left( \hbar m_\mathrm{e} \omega_0^3 /2 N_{\mathrm{e}} \right)^{1/2}$.

\section{Plasmonic bandstructure}
\label{sec3}

We start by fully characterizing the fundamental properties of the plasmonic bipartite chain in the absence of the damping mechanisms. To this end, we diagonalize exactly in the sequel the Hamiltonian \eqref{eq01} by means of bosonic Bogoliubov transformations. We provide analytic expressions for the plasmonic dispersion relation and its associated eigenstates. 

Using periodic Born-von Karman boundary conditions and introducing the bosonic operators in momentum space $a_q^{\sigma}$ and $b_q^{\sigma}$ through the Fourier transforms \cite{footnote1}
\begin{subequations}
\label{eq02}
\begin{align}
  a_n^{\sigma} &= \frac{1}{\sqrt{\mathcal{N}}} \sum_q \mathrm{e}^{\mathrm{i} n q d}\ a_q^{\sigma}, \\
  b_n^{\sigma} &= \frac{1}{\sqrt{\mathcal{N}}} \sum_q \mathrm{e}^{\mathrm{i} n q d}\ b_q^{\sigma}, 
 \end{align}
\end{subequations}
the real space Hamiltonian \eqref{eq01} becomes
\begin{align}
\label{eq03}
  H_{\mathrm{pl}} =&\;  \sum_{q \sigma} \hbar \omega_0 \left( {a_q^{\sigma}}^{\dagger} a_q^{\sigma} + {b_q^{\sigma}}^{\dagger} b_q^{\sigma}  \right) \nonumber\\
&+  \sum_{q \sigma}\left[\hbar \nu_q^{\sigma} {a_q^{\sigma}}^{\dagger}  \left( b_q^{\sigma} + {b_{-q}^{\sigma}}^{\dagger}  \right) + \mathrm{h.c.}\right],
\end{align}
where we have introduced the quantity 
\begin{equation}
\label{eq04}
 \nu_q^{\sigma} = \eta_{\sigma} \left( \Omega_1 +  \Omega_2\ \mathrm{e}^{ - \mathrm{i} q d} \right). 
\end{equation}
The quadratic Hamiltonian \eqref{eq03} is diagonalized with the bosonic Bogoliubov transformation \cite{Weick2013, Sturges2015}
\begin{equation}
\label{eq05}
 \beta_{q\tau}^{\sigma} = \cosh{\theta_{q\tau}^{\sigma}}\ \alpha_{q\tau}^{\sigma} + \sinh{\theta_{q\tau}^{\sigma}}\ {\alpha_{-q\tau}^{\sigma}}^{\dagger}, 
\end{equation}
where the index $\tau = + 1$ ($-1$) corresponds to the upper (lower) collective plasmon band. In Eq.\ \eqref{eq05}, the Bogoliubov coefficients read
\begin{subequations}
\label{eq06}
\begin{equation}
\label{eq06b}
  \cosh{\theta_{q\tau}^{\sigma}} = \frac{1}{\sqrt{2}} \left( \frac{1 + \tau |\nu_q^{\sigma}| / \omega_0 }{\sqrt{ 1 + 2 \tau |\nu_q^{\sigma}| / \omega_0}} + 1 \right)^{1/2}
\end{equation}
and 
\begin{equation}
\label{eq06c}
  \sinh{\theta_{q\tau}^{\sigma}} = \frac{\tau}{\sqrt{2}} \left( \frac{1 + \tau |\nu_q^{\sigma}| / \omega_0}{\sqrt{ 1 + 2 \tau |\nu_q^{\sigma}| / \omega_0 }} - 1 \right)^{1/2},
  \end{equation}
\end{subequations}
and the auxiliary bosonic operators are
\begin{equation}
\label{eq06a}
 \alpha_{q\tau}^{\sigma} = \frac{1}{\sqrt{2}}  \left( a_q^{\sigma} + \tau \frac{\nu_q^{\sigma}}{|\nu_q^{\sigma}|} b_q^{\sigma} \right).
\end{equation}
Thus we arrive at the diagonalized form of Eq.~\eqref{eq03} \cite{footnote2},
\begin{equation}
\label{eq07}
  H_{\mathrm{pl}} = \sum_{q \sigma \tau} \hbar \omega_{q\tau}^{\sigma} {\beta_{q\tau}^{\sigma}}^\dagger  \beta_{q\tau}^{\sigma},
\end{equation}
where the eigenfrequencies of the collective plasmonic modes are
\begin{equation}
\label{eq08}
  \omega_{q\tau}^{\sigma} = \omega_0 \sqrt{1 + 2 \tau \frac{|\nu_q^{\sigma}|}{\omega_0}  }.
\end{equation}

\begin{figure}[tb]
 \includegraphics[width=\columnwidth]{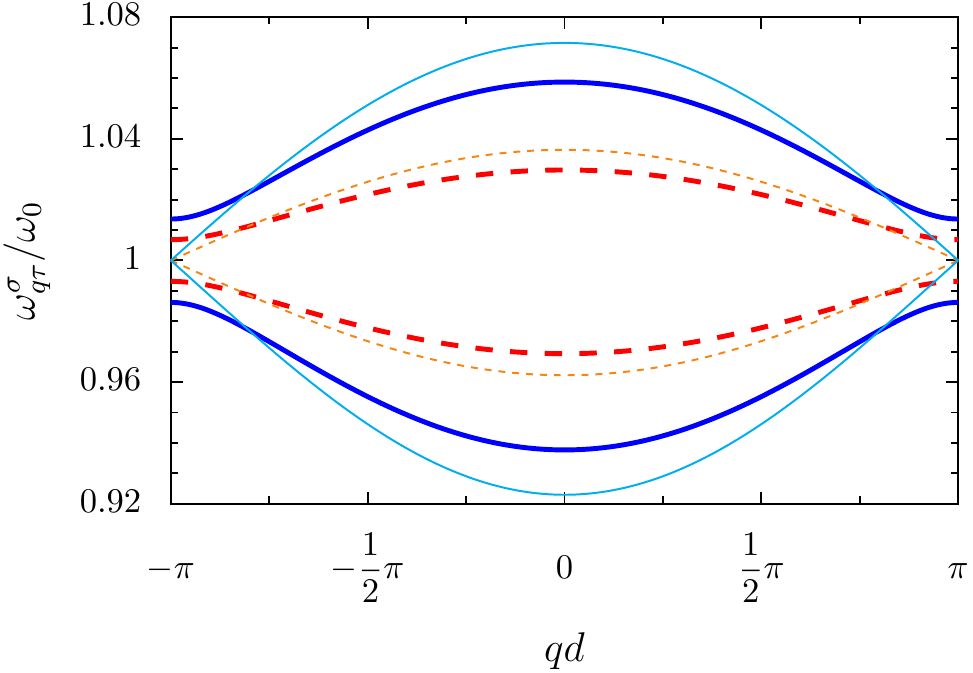}
 \caption{Collective plasmonic dispersion relation from Eq.\ \eqref{eq08} for the transverse (red and orange dashed lines, $ \sigma = x = y$) and longitudinal (blue and cyan solid lines, $ \sigma = z$) modes for both the upper ($\tau = +1$) and the lower ($\tau = -1$) bands. The undimerized case (thinner lines) with $ d_1 = d_2 = 3 a $ and a typical dimerized case (thicker lines) with $ d_1 = 3 a $ and $ d_2 = 3.5 a $ are both displayed.}
 \label{fig2}
\end{figure}

The dispersion relation \eqref{eq08} is plotted in Fig.\ \ref{fig2} for the regular chain limit where $d_1 = d_2$ (thinner lines) and a typical dimerized chain with $d_1 \neq  d_2$ (thicker lines) for both the transverse (dashed red and orange lines) and longitudinal (solid blue and cyan lines) collective plasmons. One notices an immediate analogy to the energy bands in a Peierls chain \cite{Peierls1955}, as the dimerization has the effect of opening up a small gap  of the order of $\Delta^\sigma\simeq2 |\eta_{\sigma}| |\Omega_1-\Omega_2|$ at the edges of the first Brillouin zone ($qd = \pm \pi$). 
The longitudinal collective plasmon mode has a gap which is about twice as large as compared to the transverse modes because the attractive interaction between pairs of dipoles parallel to the chain is twice as much as between dipoles perpendicular to the chain which experience a repulsive interaction [cf.\ the factor $\eta_\sigma$ appearing in Eq.\ \eqref{eq01}].

Notably, the high- (low-)energy band for the transverse polarization ($\sigma=x,y$) corresponds to a situation where the dipole moments on each 
nanoparticle within a given dimer are in-phase (out-of-phase). Conversely, for the longitudinal modes ($\sigma=z$), the high- (low-)energy band 
corresponds to out-of-phase (in-phase) dipole moments within each dimer. This has important consequences for the radiative properties of such states, as the in-phase bands [($\sigma=x,y$, $\tau=+1$) and ($\sigma=z$, $\tau=-1$)] and out-of-phase bands [($\sigma=x,y$, $\tau=-1$) and $\sigma=z$, $\tau=+1$)] have drastically different couplings to the photonic environment (see Sec.\ \ref{sec:radiative}). 

In Appendix \ref{appendA} we calculate the collective plasmon dispersion where the dipole-dipole interaction is considered beyond nearest neighbors. Such additional terms in the near-field interaction only lead to a minor correction to the dispersion relation shown in Fig.\ \ref{fig2}, justifying the nearest-neighbor approximation to the near-field interaction which we use in the remainder of this work.

\section{Massive Dirac-like collective plasmons}
\label{sec4}

Now that the plasmonic Hamiltonian \eqref{eq01} has been diagonalized in Sec.\ \ref{sec3}, we are in a position 
to study in more details the properties of its associated eigenstates. In this section we show that at the edge of the Brillouin zone, the system can effectively be described by a 1D Dirac-like Hamiltonian, which gives rise to a Klein tunneling phenomenon for the collective plasmons. We further comment on the experimental setups required to observe the aforementioned tunneling effect.

To analyze the physics occurring at the edge of the Brillouin zone, we Taylor-expand the plasmon dispersion \eqref{eq08} in the vicinity of $q d  = \pi + k d$, with $k d \ll 1$. In the regime $\Omega_{1, 2} \ll \omega_0$, we find that the collective excitations are governed by a pseudorelativistic spectrum 
\begin{equation}
\label{eq10}
  \omega_{k\tau}^{\sigma} = \omega_0 + \tau |\eta_{\sigma}| \sqrt{(\Omega_1 - \Omega_2 )^2 + \Omega_1 \Omega_2 (kd)^2},
\end{equation}
which is reminiscent of the spectrum of massive Dirac particles, $E_{\mathrm{D}} = \pm \sqrt{(mc^2)^2 + (pc)^2}$, where $\pm$ distinguishes the electron and positron particle species, and $p$ and $m$ are the momentum and mass of the particle, respectively. 

The pseudorelativistic spectrum \eqref{eq10} indicates a connection to a Dirac-like Hamiltonian, as was first noticed in the continuum quantum field theory study of the (fermionic case) of diatomic polymers \cite{Jackiw1983}. Working in the regime of first order in $\Omega_{1, 2} \ll \omega_0$, one finds from Eq.~\eqref{eq06} that $\cosh{\theta_{q\tau}^{\sigma}} \simeq 1$ and $\sinh{\theta_{q\tau}^{\sigma}} \simeq \tau |\nu_q^{\sigma}| / 2 \omega_0$, where $|\nu_q^{\sigma}| / \omega_0 \simeq |1 - \Omega_2 / \Omega_1 | \Omega_1 / \omega_0$. It follows that for small dimerizations $(\Omega_1 \simeq \Omega_2)$, $\sinh{\theta_{q\tau}^{\sigma}} \simeq 0$, such that the plasmonic Hamiltonian \eqref{eq07} can be expressed in terms of the auxiliary operators of Eq.\ \eqref{eq06a}, 
$H_{\mathrm{pl}} = \sum_{k \sigma \tau} \hbar \omega_{k\tau}^{\sigma} {\alpha_{k\tau}^{\sigma}}^{\dagger} \alpha_{k\tau}^{\sigma}$.
Performing the summation over $\tau$, we obtain a $2 \times 2$ matrix Hamiltonian 
$H_{\mathrm{pl}} = \sum_{k \sigma} {\Psi_{k}^{\sigma}}^{\dagger} \mathcal{H}_{k}^{\sigma} \Psi_k^{\sigma}$, with 
\begin{equation}
\label{eq12}
\mathcal{H}_{k}^{\sigma} = \hbar 
\begin{pmatrix}
  \omega_0 & {\nu_k^{\sigma}}^* \\
  \nu_k^{\sigma} & \omega_0 
 \end{pmatrix} 
,
\end{equation}
where $\nu_k^{\sigma} = \eta_{\sigma} \left( \Omega_1 - \Omega_2 + \mathrm{i}\, \Omega_2 k d \right)$ and where $\Psi_k^{\sigma} = (a_k^{\sigma}, b_k^{\sigma})$. The associated eigenvectors are in the form of a Dirac spinor
\begin{equation}
\label{eq13}
   \ket{\psi_{k\zeta}^{\sigma}} = \frac{1}{\sqrt{2}}\left(
 \begin{array}{c}
 1 \\ \zeta \tfrac{\nu_k^{\sigma}}{|\nu_k^{\sigma}|}
 \end{array}
\right), 
\end{equation}
which is reminiscent of eigenvectors found in 1D Dirac materials such as single-wall carbon nanotubes \cite{Kane1997}. The index $\zeta$ denotes whether the excitation can be thought of (in the language of semiconductor physics) as hole-like ($\zeta=-1$) or electron-like ($\zeta=+1$), as is the case with the spinor of graphene \cite{Neto2009}. 

For the Dirac-like nature of the Hamiltonian \eqref{eq12} to become apparent, we utilize the unitary transform $\mathcal{U} = (\sigma_x + \sigma_z)/\sqrt{2}$, to reach $\bar{\mathcal{H}}_{k}^{\sigma}  = \mathcal{U} \mathcal{H}_{k}^{\sigma} \mathcal{U}^{\dagger}$, with  
 \begin{equation}
\label{eq14}
   \bar{\mathcal{H}}_{k}^{\sigma} = \hbar \omega_0 \mathbbm{1}_2 + \eta_{\sigma} \hbar  [ \left( \Omega_1 - \Omega_2 \right) \sigma_z - \Omega_2  k d \sigma_y ].
\end{equation}
Here, $\mathbbm{1}_2$ is the $2 \times 2$ identity matrix and $\sigma_x$, $\sigma_y$, and $\sigma_z$ are Pauli's spin matrices. The parallels with the 1D form of the Dirac Hamiltonian $\mathcal{H}_{\mathrm{D}} = m c^2 \sigma_z + c p \sigma_y$ are evident, up to the unimportant constant energy shift $\hbar\omega_0 \mathbbm{1}_2$. 
A striking feature in the effective Hamiltonian \eqref{eq14} is that the role of the mass term is played by a quantity 
$\propto (\Omega_1 - \Omega_2)$ which can be modified in both magnitude and sign through the interparticle distances $d_1$ and $d_2$ [see Eq.\ \eqref{eq:Omega}]. This tunability has significant implications for both pseudorelativistic Klein tunneling (see Appendix \ref{appendB}) and the topological properties of bipartite metallic nanoparticle chains (cf.\ Sec.\ \ref{sec5}). The present realization of 1D Dirac-like bosons encapsulated in the Hamiltonian \eqref{eq14} complements the recent discovery of their counterparts in 2D honeycomb metasurfaces \cite{Weick2013, Sturges2015, Banerjee2016}.

A prominent feature of excitations described by a Dirac equation is their ability to be transmitted through a potential barrier with highly suppressed backscattering \cite{Klein1929}, the so-called Klein paradox. For the system of nanoparticles we investigate here, a plasmonic version of Klein tunneling can be realized by arranging a dimerized chain into two distinct sections, which we distinguish with the notation $\mathrm{L}$ (for left) and $\mathrm{R}$ (for right). These two sections have different resonance frequencies $\omega_0^{\mathrm{L, R}}$ to mimic the application of an electrostatic potential step, as can be inferred from the first term in the right-hand side of the Dirac-like Hamiltonian \eqref{eq14} (see Appendix \ref{appendB} for a detailed calculation of the corresponding transmission probability).
Such a scenario may be experimentally achieved by the three following different means: (i) using two different metals (for example silver and gold); (ii) modifying the electrostatic environment of one chain with respect to the other, using an embedding dielectric medium; or (iii) using nanoparticles of different sizes, for which quantum-size effects such as the spill-out of the electronic density outside the nanoparticle \cite{Kreibig1995, Brack1993} as well as coupling to the electronic environment \cite{Weick2006} influence the position of the surface plasmon resonance.

\section{Topological properties of the collective plasmons}
\label{sec5}

The appearance of a Dirac equation with its associated spinor wavefunctions in the description of the collective plasmons in the dimerized chain (see Sec.\ \ref{sec4}) indicates that a nontrivial Berry phase may arise in this system. In this section we study the implications of such physics on bosonic quasiparticles, classify the topological properties which are found to arise and investigate the existence of edge states. As a result, we uncover our system as a `plasmonic topological insulator', which has the ability to both harbor and exploit topologically-protected midgap states.

\subsection{Zak phase}
\label{sec:Zak}

The Berry phase \cite{Berry1984} when applied to the dynamics of electrons in periodic solids in 1D is known as the Zak phase \cite{Zak1989}. For our purposes here, its utility arises because it is related to the existence of edge states, in a manifestation of a bulk-edge correspondence principle \cite{Delplace2011}. The Zak phase is defined as an integration over the first Brillouin zone of the Berry connection $\mathcal{A}=- \mathrm{i} \bra{\psi_{q\tau}^{\sigma}} \partial_{q} \ket{\psi_{q\tau}^{\sigma}}$, explicitly
\begin{equation}
\label{eq19}
\vartheta_{\mathrm{Z}} = \int_{\mathrm{1st\ BZ}} \mathrm{d} q \, \mathcal{A}, 
\end{equation}
which requires knowledge of the plasmonic eigenstates $\ket{\psi_{q\tau}^{\sigma}}$. We may write the plasmonic eigenstates via identification with the Bogoliubov coefficients of Eq.~\eqref{eq05}, namely 
\begin{equation}
\label{eq20}
 \ket{\psi_{q\tau}^{\sigma}} = \frac{1}{\sqrt{2}}
 \begin{pmatrix}
 \cosh{\theta_{q\tau}^{\sigma}} \\ 
 \tau \frac{\nu_q^{\sigma}}{|\nu_q^{\sigma}|} \cosh{\theta_{q\tau}^{\sigma}} \\ 
 \sinh{\theta_{q\tau}^{\sigma}} \\ 
 \tau \frac{\nu_q^{\sigma}}{|\nu_q^{\sigma}|} \sinh{\theta_{q\tau}^{\sigma}}
 \end{pmatrix}
\end{equation}
in the basis $( a_q^{\sigma}, b_q^{\sigma}, a_{-q}^{\sigma \dagger}, b_{-q}^{\sigma \dagger})$ \cite{footnote3}. Then, proceeding to calculate the Zak phase from Eq.~\eqref{eq19} with Eq.~\eqref{eq20}, we find
\begin{equation}
\label{eq21}
\vartheta_{\mathrm{Z}} =
\begin{cases} 
   0, \quad \Omega_1 > \Omega_2 & (d_1 < d_2) \\
   \pi, \quad \Omega_1 < \Omega_2 & (d_1 > d_2)
  \end{cases}
\end{equation}
defined modulo $2\pi$ (since the Berry connection $\mathcal{A}$ is not gauge invariant). Thus the bipartite chain has a trivial Zak phase $\vartheta_{\mathrm{Z}} = 0$ when $d_1<d_2$ [$\Omega_1>\Omega_2$, cf.\ Eq.\ \eqref{eq:Omega}] and a nontrivial one $\vartheta_{\mathrm{Z}} = \pi$ when $d_1>d_2$  ($\Omega_1<\Omega_2$), with a topological phase transition at the crossover point of the regular chain $d_1 = d_2$, where the bandgap closes (see Fig.\ \ref{fig2}). Equation~\eqref{eq21} alludes to the presence of highly confined edge states in the nontrivial phase, corresponding to topologically-protected localized collective plasmons, which we investigate in detail in the remainder of this section.

\subsection{Plasmonic edge states}
\label{sec:edge_states}

\begin{figure}[tb]
 \includegraphics[width=1.0\columnwidth]{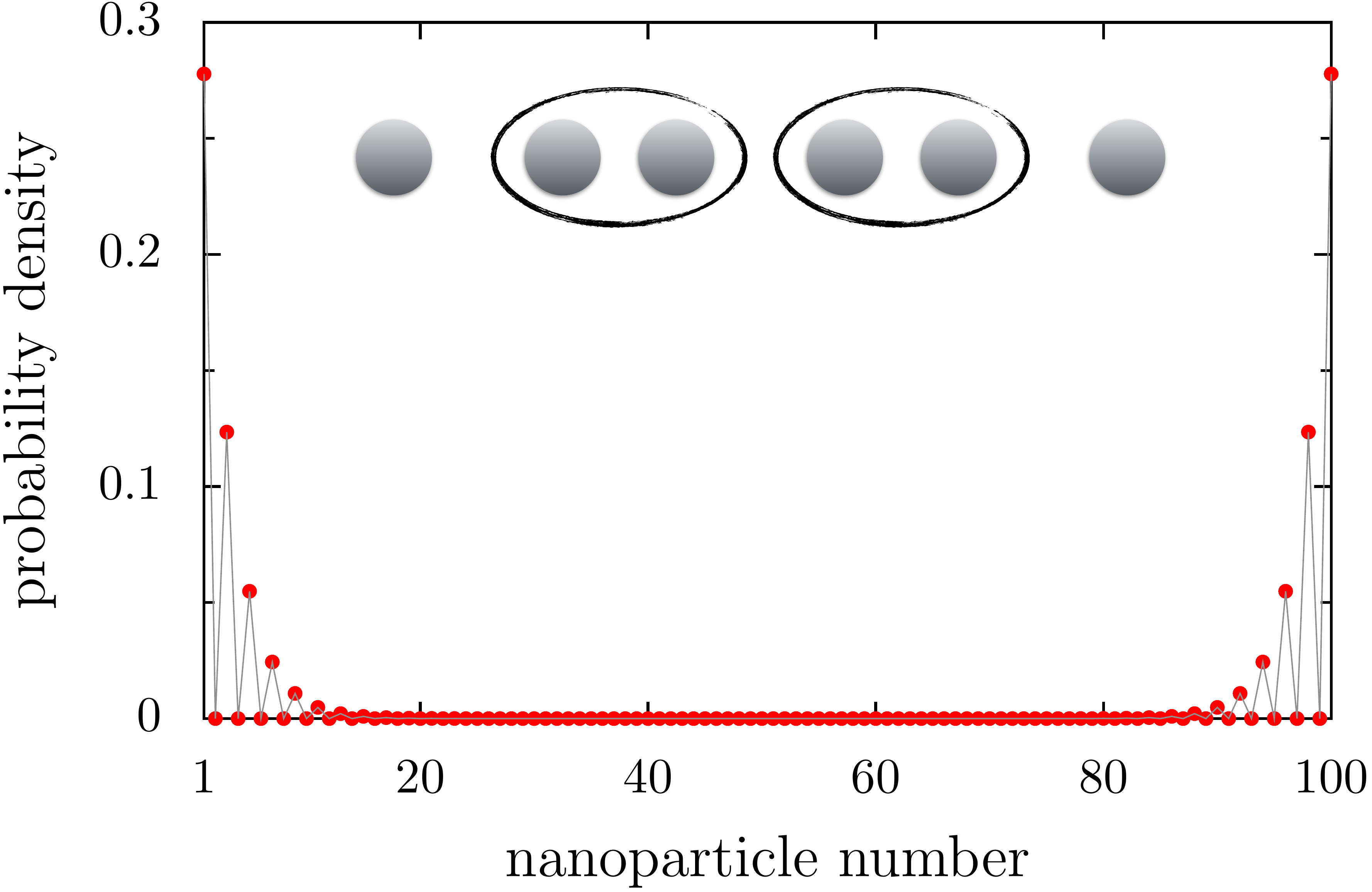}
 \caption{Plasmonic probability density corresponding to the midgap state at frequency $\omega_0$ (red dots, thin gray line: guide for the eye) at each site in a chain of $100$ nanoparticles, where $\Omega_1 / \Omega_2 = 2/3$. Inset: Sketch of a typical chain comprised of an even number of nanoparticles, where the loops depict the favored dimerization.}
 \label{fig3}
\end{figure}

We now search for a midgap collective plasmon state at the frequency $\omega_0$ supported by the finite bipartite chain described by the Hamiltonian \eqref{eq01} and composed of an even number of nanoparticles (for completeness, we consider in Appendix \ref{appendD} the case where the chain is composed of an odd number of constituents). To this end, we work in real space with open boundary conditions. Due to the requirement of preserving bosonic statistics, diagonalization of Eq.~\eqref{eq01} requires the framework of paraunitary transforms \cite{footnote5}. The result of this numerical procedure is shown in Fig.\ \ref{fig3}, where we plot the probability density associated with the midgap state at each nanoparticle site in the regime $\Omega_1 < \Omega_2$ ($d_1>d_2$). As foreshadowed by the nontrivial Zak phase $\vartheta_\mathrm{Z}=\pi$ found in Sec.\ \ref{sec:Zak} in that case, both edges are highly plasmonically active. We have also checked numerically that, as expected, there is no midgap state when $\Omega_1 > \Omega_2$ ($d_1<d_2$) for which one finds a vanishing Zak phase [cf.\ Eq.\ \eqref{eq21}].

To better understand the critical factor determining the formation of the midgap edge states, we neglect the nonresonant terms in Eq.~\eqref{eq01} and maintain only the regular `hopping' terms in the following analytic analysis. This rotating wave approximation (RWA) is justified post hoc due to the excellent comparison to the results found via full numerical diagonalization of the Hamiltonian \eqref{eq01} in the limit of a large number of nanoparticles. We thus start from the real space Hamiltonian \eqref{eq01} in the form $H_{\mathrm{pl}} = \sum_{\sigma} H_{\mathrm{pl}}^{\sigma}$, with 
\begin{align}
  H_{\mathrm{pl}}^\sigma \simeq&\; \hbar \omega_0 \sum_{n=1}^\mathcal{N} 
  \left( {a_n^{\sigma}}^{\dagger} a_n^{\sigma} + {b_n^{\sigma}}^{\dagger}   b_n^{\sigma} \right) \nonumber\\
&+ \hbar \Omega_1  \eta_{\sigma} \sum_{n=1}^\mathcal{N} \left( {a_n^{\sigma}}^{\dagger}b_n^{\sigma}+\mathrm{h.c.} \right) \nonumber \\
  &+ \hbar \Omega_2  \eta_{\sigma} \sum_{n=1}^{\mathcal{N}-1 } \left( {a_{n+1}^{\sigma}}^{\dagger}   b_n^{\sigma} + \mathrm{h.c.}  \right). 
\end{align}
We write the above Hamiltonian as 
$H_{\mathrm{pl}}^{\sigma} =  {\Phi^{\sigma}}^{\dagger} {\sf{H}}_\mathrm{pl}^{\sigma} \Phi^{\sigma}$ where $\Phi^{\sigma} = (a_1^{\sigma}, b_1^{\sigma}, \ldots, a_{\mathcal{N}}^{\sigma}, b_{\mathcal{N}}^{\sigma})$, and where $\sf{H}_\mathrm{pl}^{\sigma}$ is an even, symmetric, tridiagonal matrix. The diagonal elements of $\sf{H}_\mathrm{pl}^{\sigma}$ are all equal to $\omega_0$, while the off-diagonal elements alternate between $\eta_{\sigma} \Omega_1$ and $\eta_{\sigma} \Omega_2$. Diagonalizing the matrix $\sf{H}_\mathrm{pl}^{\sigma}$ amounts to the following system of equations for the coefficients $(A_1^\sigma, B_1^\sigma, \ldots, A_\mathcal{N}^\sigma, B_\mathcal{N}^\sigma)$ associated with the operator $\Phi^{\sigma}$:
\begin{subequations}
\label{eqbig1}
  \begin{equation}
\label{eqbig2}
   \left( \omega_0 - \omega \right) B_{n}^\sigma + \eta_{\sigma} \Omega_1 A_{n}^\sigma + 
   \eta_{\sigma} \Omega_2 A_{n+1}^\sigma = 0, \quad n \in [1, \mathcal{N} - 1]
\end{equation}
 \begin{equation}
\label{eqbig3}
   \left( \omega_0 - \omega \right) A_{n}^\sigma + \eta_{\sigma} \Omega_1 B^\sigma_{n-1} 
   + \eta_{\sigma} \Omega_2 B^\sigma_{n} = 0, \quad n \in [2, \mathcal{N}]
\end{equation}
 \begin{equation}
\label{eqbig4}
    \left( \omega_0 - \omega \right) A_{1}^\sigma + \eta_{\sigma} \Omega_1 B^\sigma_{1} = 0, 
\end{equation}
 \begin{equation}
 \label{eqbbig4}
\left( \omega_0 - \omega \right) B_{\mathcal{N}}^\sigma + \eta_{\sigma} \Omega_1 A^\sigma_{\mathcal{N}} = 0.
\end{equation}
\end{subequations}
Searching for midgap states at $\omega_0$ produces from Eqs.~\eqref{eqbig4} and \eqref{eqbbig4} the boundary conditions $B_1^\sigma = A^\sigma_{\mathcal{N}} = 0$ at the chain edges. 
In the bulk, the recurrence relations formed from Eqs.\ \eqref{eqbig2} and \eqref{eqbig3} asymptotically satisfy these boundary conditions as long as $\Omega_1 < \Omega_2$ ($d_1>d_2$), with the probability densities on each site given by 
\begin{subequations}
\label{eq88}
\begin{align}
  \left|A_n^\sigma\right|^2 &= \frac{(\Omega_1/\Omega_2)^2-1}{2\left[(\Omega_1/\Omega_2)^{2\mathcal{N}}-1\right]}\left( \frac{\Omega_1}{\Omega_2} \right)^{2 (n-1)}, \\
  \left|B^\sigma_{n}\right|^2 &= \frac{(\Omega_1/\Omega_2)^2-1}{2\left[(\Omega_1/\Omega_2)^{2\mathcal{N}}-1\right]}\left( \frac{\Omega_1}{\Omega_2} \right)^{2 (\mathcal{N}-n)},
\end{align}
\end{subequations}
with $n \in [1, \mathcal{N}]$. A plot of the above expressions is indistinguishable from the one of the full numerical solution in Fig.\ \ref{fig3} for long-enough chains  and so is not displayed in the figure \cite{footnote:length}. The exponential  behavior of the plasmonic probability density associated with the midgap state at frequency $\omega_0$ can be understood with the help of the simple picture in the inset of Fig.\ \ref{fig3}, where we sketch a representative chain consisting of an even number of nanoparticles. The $\Omega_1 < \Omega_2$ dimerization favors a pairing-up of nanoparticles as depicted with the drawn loops, leaving seemingly isolated nanoparticles at both ends of the chain which are both free to harbor the majority of the plasmonic activity.

\subsection{Topologically-protected Jackiw-Rebbi-like state}

Another edge state may form at the interface between two topologically-distinct bipartite chains (both considered for simplicity to have the same resonance frequency $\omega_0$). At the formed `domain wall', a topologically-protected kink-like bound state arises at the midgap frequency $\omega_0$. In this way, the state is reminiscent of a Jackiw-Rebbi mode \cite{Jackiw1976}, an exponentially pinned eigenstate which arises at the boundary where the mass term changes sign in a Dirac equation with position-dependent mass. 

Let us consider bringing together two bipartite chains, identified by their different Zak phases $\vartheta_{\mathrm{Z}}=0$ and $\pi$, and characterized by their coupling constants
\begin{equation}
\label{eq:OmegaSoliton}
\Omega_{1, 2}^{\vartheta_{\mathrm{Z}}} = \frac{\omega_0}{2} \left(\frac{a}{d_{1, 2}^{\vartheta_{\mathrm{Z}}}}\right)^3.
\end{equation}
We choose the left-hand side chain to be topologically trivial $(\vartheta_{\mathrm{Z}} = 0, \Omega_1^0 > \Omega_2^0)$ and the right-hand side chain to be topologically nontrivial $(\vartheta_{\mathrm{Z}} = \pi, \Omega_1^{\pi} < \Omega_2^{\pi})$. Furthermore, the connection of these two chains is described with the interface coupling constant $\Omega_{\mathrm{i}}$. For simplicity, we investigate the case of an odd total number of nanoparticles, comprised of $2 \mathcal{N}$ nanoparticles in the left-hand side chain and $2 \mathcal{N} + 1$ nanoparticles in the right-hand side chain. The resulting real space Hamiltonian can be written down in analogy to the Hamiltonian \eqref{eq01}.

\begin{figure}[tb]
 \includegraphics[width=1.0\columnwidth]{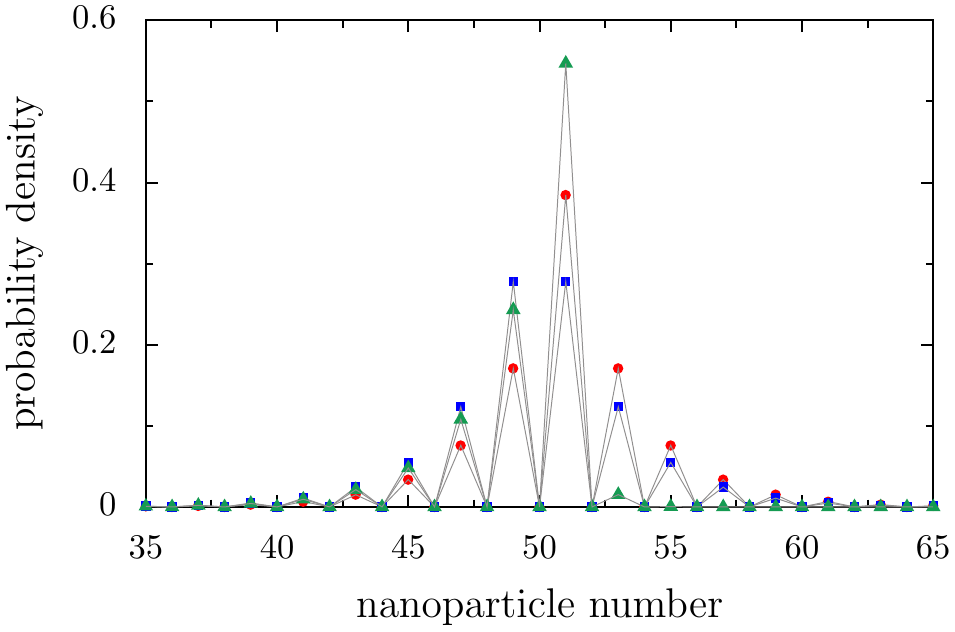}
 \caption{Plasmonic probability density corresponding to the midgap state at frequency $\omega_0$ at each site in a chain of $101$ nanoparticles, for three different choices of the coupling constants. In the two symmetric setups $\Omega_2^0 / \Omega_1^0 = \Omega_1^{\pi} / \Omega_2^{\pi} = 2/3$, with the interface coupling constant being either $\Omega_{\mathrm{i}} = \Omega_2^0$ (red dots) or $\Omega_{\mathrm{i}} = \Omega_1^0$ (blue squares). In the asymmetric setup $\Omega_2^0 / \Omega_1^0 = 2/3$, $\Omega_1^{\pi} / \Omega_2^{\pi} = 1/6$, and $\Omega_{\mathrm{i}} = \Omega_2^0$ (green triangles). The thin gray lines are guides for the eye.}
 \label{fig4}
\end{figure}

As alluded to in Sec.\ \ref{sec:edge_states}, one may diagonalize quadratic, bosonic Hamiltonians by employing paraunitary transforms \cite{Tsallis1978, Colpa1978, Kawaguchi2012}. Following this approach, we obtain the probability density of the midgap state at frequency $\omega_0$ as a function of the nanoparticle site, which is plotted in Fig.~\ref{fig4}. We consider three fundamental varieties of edge states which can be obtained by expedient choices of internanoparticle separations. Firstly, we consider symmetric states with coupling constants $\Omega_2^0 / \Omega_1^0 = \Omega_1^{\pi} / \Omega_2^{\pi} = 2/3$. With the interface coupling constant $\Omega_{\mathrm{i}} = \Omega_2^0$ (red dots) the edge state is single-peaked and centered on the nanoparticle corresponding to the $A$ sublattice in the $(\mathcal{N}+1)$st dimer; while for $\Omega_{\mathrm{i}} = \Omega_1^0$ (blue squares) the state is double-peaked, with equal maxima on the nanoparticles corresponding to the $A$ sublattice in the ${\mathcal{N}}$th and $({\mathcal{N}+1})$st dimer. Finally, a typical asymmetric midgap state is shown in Fig.\ \ref{fig4} with the parameters $\Omega_2^0 / \Omega_1^0 = 2/3$, $\Omega_1^{\pi} / \Omega_2^{\pi} = 1/6$, and $\Omega_{\mathrm{i}} = \Omega_2^0$ (green triangles). In this case, the significant asymmetry causes the midgap state to be highly confined to the nanoparticle in sublattice $A$ in the $({\mathcal{N}+1})$st dimer. Notably, due to the experimental possibility to position metallic nanoparticles in any desired location, the system provides some tunability over the size of the localization length of the Jackiw-Rebbi-like state. 

The topologically-protected midgap states above can be well described using the RWA, which disregard the nonresonant terms which appear in the governing Hamiltonian. The validity of this conjecture is, as in Sec.\ \ref{sec:edge_states}, proved after the fact by comparing the analytic expressions obtained in the RWA with the outcome of the numerical diagonalization of the full Hamiltonian. Such an RWA analysis proceeds in a similar manner to that carried out in Sec.\ \ref{sec:edge_states}. One finds the complete extinction of plasmonic activity on the $B$ sublattice, $\left|B_{n}^\sigma\right|^2 = 0$, with $n \in [1, 2 \mathcal{N}]$. 
This is a consequence of the chiral (or sublattice) symmetry of the bipartite chain \cite{Asboth}.
Meanwhile on the $A$ sublattice, we find an exponentially-pinned probability density profile
\begin{subequations}
\label{eq:ProbSoliton}
\begin{equation}
  \left|A_n^\sigma\right|^2 = \left|A_{\mathcal{N}}^\sigma\right|^2 \left( \frac{\Omega_2^0}{\Omega_1^0} \right)^{2 (\mathcal{N}-n)} 
\end{equation}
for $n \in [1, \mathcal{N} - 1]$ and 
\begin{equation}
  \left|A_n^\sigma\right|^2 = \left|A_{\mathcal{N}}^\sigma\right|^2 
  \left( \frac{\Omega_1^{0}}{\Omega_\mathrm{i}} \right)^2
  \left( \frac{\Omega_1^{\pi}}{\Omega_2^{\pi}} \right)^{2 (n-\mathcal{N}-1)}
\end{equation}
\end{subequations}
for $n \in [\mathcal{N} + 1, 2 \mathcal{N} + 1]$
in terms of the probability density on the final $A$ site of the left-hand side chain,
\begin{align}
\label{eq:ProbSoliton2}
\left|A_{\mathcal{N}}^\sigma\right|^2 =&
 \left[ \left( \frac{\Omega_2^0}{\Omega_1^0} \right)^2 \frac{\left( \Omega_2^0 / \Omega_1^0  \right)^{2 (\mathcal{N}-1)} - 1}{\left( \Omega_2^0 / \Omega_1^0  \right)^2 - 1}
\right.\nonumber\\
& + \left. 1 + \left( \frac{\Omega_1^0}{\Omega_{\mathrm{i}}} \right)^2 \frac{\left( \Omega_1^{\pi} / \Omega_2^{\pi}  \right)^{2 (\mathcal{N}+1)} - 1}{\left( \Omega_1^{\pi} / \Omega_2^{\pi}  \right)^2 - 1}  \right]^{-1}. 
\end{align}
Since these analytic expressions are imperceptible from the numerically exact solution displayed in Fig.~\ref{fig4}, they are not plotted in the figure. Equation \eqref{eq:ProbSoliton} defines a bosonic version of a strongly-localized midgap state. Its existence at the domain-wall boundary only requires the connection of two chains of different Zak phases. The precise details of the chain are unimportant, so it corresponds to a topologically-protected plasmonic bound state. 

The analysis above provides a physical realization of various bosonic Jackiw-Rebbi-like states, and gives rise to the chance of witnessing such exotic states experimentally, especially when compared to fermionic systems such as conducting polymers where such an observation is a challenging task.

\section{Lifetime of the collective modes}
\label{sec6}

Now that the topological properties of the purely plasmonic problem encapsulated in the Hamiltonian \eqref{eq01} have been discussed, here we analyze in detail the influence of both the photonic and electronic environments [Eqs.\ \eqref{eq00343401} and \eqref{eq0009}, respectively] onto plasmon lifetimes. 
In particular, we calculate the radiative and size-dependent nonradiative (Landau damping) losses encountered by the collective plasmons in bipartite chains of metallic nanoparticles, 
and comment on the various challenges to experimentally observe their topological properties.

\subsection{Radiation damping}
\label{sec:radiative}

The plasmonic system sketched in Fig.\ \ref{fig1} is subject to radiative losses, which arise due to the coupling of the collective plasmons along the nanoparticle chain to vacuum electromagnetic field modes described by Eq.\ \eqref{eq00343401} and entering the system Hamiltonian \eqref{eq:H}. Treating the plasmon-photon coupling \eqref{eq0003} as a perturbation, the Fermi golden rule result for the radiative decay rate of the collective plasmon with momentum $q$ in the branch $\tau$ and polarization $\sigma$ reads
\begin{equation}
\label{eq0000000004}
\gamma_{q\tau}^{\sigma, \mathrm{r}} = \frac{2 \pi}{\hbar^2} \sum_{\mathbf{k}, \hat{\lambda}_{\mathbf{k}}} 
\left|\bra{0_{q\tau}^{\sigma}, 1_{\mathbf{k}}^{\hat{\lambda}_{\mathbf{k}}}} H_{\mathrm{pl}\textrm{-}\mathrm{ph}} \ket{1_{q\tau}^{\sigma}, \mathrm{vac}}\right|^2 \delta (\omega_{\mathbf{k}} - \omega_{q\tau}^{\sigma}).
\end{equation}
The above expression describes the transitions between the plasmonic excited state $|1_{q\tau}^\sigma\rangle$ and the 
ground state $|0_{q\tau}^\sigma\rangle$ by spontaneous emission of photons from the electromagnetic vacuum ($|\mathrm{vac}\rangle\rightarrow|1_{\mathbf{k}}^{\hat{\lambda}_{\mathbf{k}}}\rangle$). Using Eq.\ \eqref{eq0003} together with the Bogoliubov transformation \eqref{eq05}, we arrive at 
\begin{align}
\label{eq0006}
 \gamma_{q\tau}^{\sigma, \mathrm{r}} =&\; 
 \frac{\pi^2 \omega_0^2 a^3}{\mathcal{V}} \sum_{\mathbf{k}, \hat{\lambda}_{\mathbf{k}}} \left| {\hat{\sigma}} \cdot \hat{\lambda}_{\mathbf{k}} \right|^2 
   \frac{\omega_{q\tau}^{\sigma}}{\omega_{\mathbf{k}}} 
   \nonumber\\
   &\times  \left| 1 + \tau \mathrm{e}^{- \mathrm{i} k_z d_1} \frac{{\nu_q^{\sigma}}^*}{|\nu_q^{\sigma}|} \right|^2 \left|F_{\mathbf{k}, q}\right|^2 \delta (\omega_{\mathbf{k}} - \omega_{q\tau}^{\sigma}), 
\end{align}
where the array factor is defined as 
\begin{equation}
\label{eq0004}
 F_{\mathbf{k}, q} = \frac{1}{\sqrt{\mathcal{N}}} \sum_{n=1}^\mathcal{N} \mathrm{e}^{\mathrm{i} [q n  - k_z (n-1) ]d},
\end{equation}
with $k_z$ being the $z$ component of the photon wavenumber $\mathbf{k}$.
Equation \eqref{eq0004} can easily be evaluated, yielding 
\begin{equation}
\left| F_{\mathbf{k}, q} \right|^2=\frac{1}{\mathcal{N}}\frac{\sin^2{(\mathcal{N}[q-k_z]d/2)}}{\sin^2{([q-k_z]d/2)}}, 
\end{equation}
which, in the large-chain limit ($\mathcal{N}\gg1$), simplifies to 
$
\left| F_{\mathbf{k}, q} \right|^2\simeq 2 \pi \delta \left( [ q - k_z ] d \right)
$.
The above approximation has been proved to be accurate in the case of a regular chain composed of only $20$ nanoparticles (cf.\ Fig.\ 3 in Ref.\ \cite{Brandstetter2016}). 
Using that $\sum_{\hat{\lambda}_{\mathbf{k}}}|{\hat{\sigma}} \cdot \hat{\lambda}_{\mathbf{k}}|^2=1-(\hat{\sigma}\cdot \hat{k})^2$, working in the continuum limit ($\mathcal{V}\rightarrow\infty$), and replacing the sum over $\mathbf{k}$ by a three-dimensional integral in Eq.\ \eqref{eq0006}, we finally arrive at the result
\begin{align}
\label{eq0008}
 \gamma_{q\tau}^{\sigma, \mathrm{r}} =&\; \frac{3 \pi |\eta_{\sigma}|\gamma_0^\mathrm{r}}{4q_0 d}  \frac{\omega_{q\tau}^{\sigma}}{\omega_0}  
 \left[  1 + \sgn{\{\eta_{\sigma}\}}  \left( \frac{c q}{\omega_{q\tau}^{\sigma}} \right)^2  \right] 
 \nonumber\\
&\times \left[ 1 + \tau \sgn{\{\eta_{\sigma}\}} \frac{\Omega_1 \cos (q d_1) + \Omega_2 \cos (q d_2)}{\sqrt{\Omega_1^2 + \Omega_2^2 + 2 \Omega_1 \Omega_2 \cos (q d)}} \right] \nonumber\\
& \times \Theta \left( \omega_q^{\sigma \tau} - c |q| \right)
\end{align}
for the radiation damping of the collective plasmons 
in terms of the single-particle result $\gamma_0^r = 2 \omega_0^4 a^3 / 3 c^3$, where $q_0=\omega_0/c$.

\begin{figure}[tb]
\includegraphics[width=1.0\columnwidth]{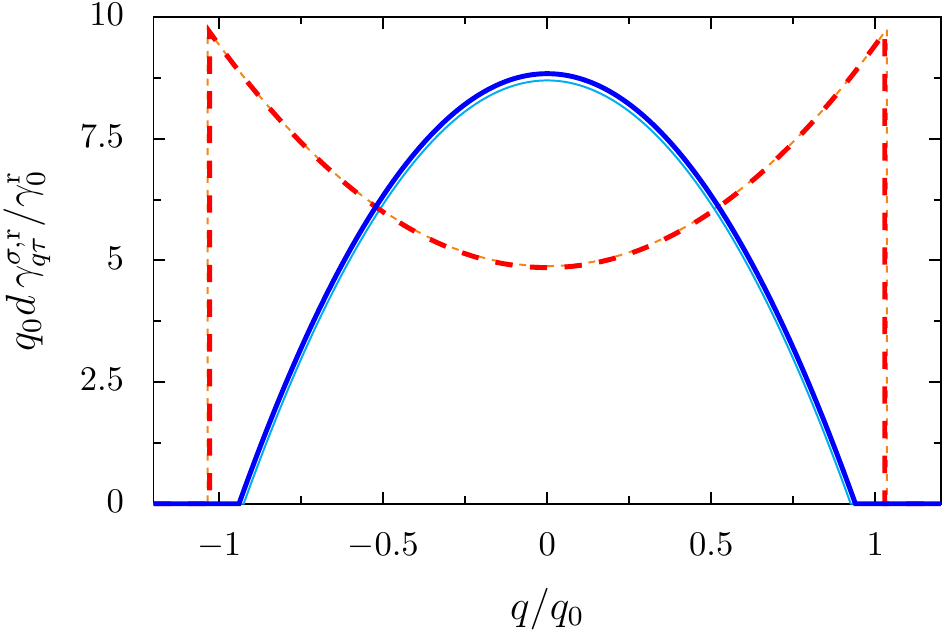}
 \caption{Radiation damping decay rate from Eq.~\eqref{eq0008} as a function of momentum for the transverse (red and orange dashed lines, $ \sigma = x = y$) and longitudinal (blue and cyan solid lines, $ \sigma = z$) collective plasmonic modes for $q_0 d = 0.5$. The undimerized case (thinner lines) with $ d_1 = d_2 = 3 a $ and a typical dimerized case (thicker lines) with $ d_1 = 3.5 a $ and $ d_2 = 3 a $ are both displayed. Only the results for the in-phase bands, i.e., the lower $\tau = -1$ band for the longitudinal polarization and the upper $\tau = +1$ band for the transverse modes, are displayed in the figure as the out-of-phase bands provide vanishingly small contributions.}
 \label{fig5}
\end{figure}

We plot in Fig.\ \ref{fig5} the radiation damping decay rate \eqref{eq0008} for the in-phase bands. The results for the regular-chain limit with $d_1 = d_2$ (thinner lines) and a typical dimerized chain with $d_1\neq d_2$ (thicker lines) are both plotted, each for the transverse (dashed red/orange lines) and longitudinal (solid blue/cyan lines) polarizations. Notably, even when the chain is in the dimerized regime, there remains only one band per polarization that provides a non-negligible contribution to the decay rate. Namely, the out-of-phase bands, corresponding to the 
longitudinal modes in the upper $\tau = +1$ band and the transverse modes in the lower $\tau = -1$ band are weakly coupled to light and are thus referred to as `dark bands'. These polarization-dependent low-loss bands will inevitably have an increased plasmonic lifetime due to being protected against radiation damping. Such dark modes arise due to destructive interferences in the far field. In contrast, the in-phase modes (the lower $\tau = -1$ band for the longitudinal polarization and the upper $\tau = +1$ band for the transverse one) within the light cone 
are subject to radiation damping. Indeed, outside of the light cone $(q \gtrsim q_0)$ the decay rate is vanishing, while inside the light cone $(q \lesssim q_0)$ the modes are both bright and superradiant, with a magnitude significantly larger than that of a single nanoparticle. 
It should also be noted that, within our model of near-field coupled nanoparticles, the topologically-protected edge states discussed in Sec.\ \ref{sec5} 
sit outside of the light cone, and hence are not subject to radiation damping. 
As can be seen in Fig.\ \ref{fig5}, dimerization has a weak effect on radiation damping, as the latter follows qualitatively the $q$-dependence of the 
simple-chain result \cite{Brandstetter2016}.

\subsection{Landau damping}
\label{sec:Landau}

The coupling \eqref{eq0010} of the collective plasmons to a bath of electron-hole pairs, described by the Hamiltonian \eqref{eq0009} and appearing in the system Hamiltonian \eqref{eq:H}, gives rise to a size-dependent nonradiative decay of a purely quantum mechanical origin: Landau damping \cite{Kawabata1966, Yannouleas1992, Weick2005}. Assuming that the plasmon-electron-hole Hamiltonian \eqref{eq0010} acts perturbatively, one can write down the following Fermi golden rule for the nonradiative decay rate of the collective plasmon with momentum $q$ in the branch $\tau$ and with the polarization $\sigma$,
\begin{align}
\label{eq00176761}
\gamma_{q\tau}^{\sigma, \mathrm{L}} =&\; 
\frac{2 \pi}{\hbar}
\sum_{n=1}^\mathcal{N}\sum_{s=A,B}\sum_{eh_{ns}}
 \left|\bra{0_{q\tau}^{\sigma}, eh_{ns} } H_{\mathrm{pl}\textrm{-}\mathrm{eh}} \ket{1_{q\tau}^{\sigma}, \mathrm{gs}_{ns} }\right|^2 
 \nonumber\\
&\times \delta (\hbar\omega_{q\tau}^{\sigma} - \varepsilon_{nse}+\varepsilon_{nsh}), 
\end{align}
with $eh_{ns}$ denoting electron-hole states in the nanoparticle belonging to the dimer $n$ and the sublattice $s$.
The above expression describes the transitions between the plasmonic excited and ground states 
by nonradiative emission of electron-hole pairs ($|\mathrm{gs}_{ns}\rangle\rightarrow|eh_{ns}\rangle$). 
Evaluating Eq.\ \eqref{eq00176761} for arbitrary $\mathcal{N}$ with the help of the Bogoliubov transformation \eqref{eq05} yields 
\begin{equation}
\label{eq0011}
 \gamma_{q\tau}^{\sigma, \mathrm{L}} = \frac{\omega_0}{\omega_{q\tau}^{\sigma}} \Sigma^{\sigma} \left( \omega_{q\tau}^{\sigma} \right)
\end{equation}
in terms of the summation over electron and hole states (assumed to be the same in each nanoparticle)
\begin{equation}
\label{eq0012}
 \Sigma^{\sigma} \left( \omega\right) = \frac{2 \pi }{\hbar} \Lambda^2\sum_{eh} | \bra{e} \sigma \ket{h} |^2 \delta \left(\hbar\omega - \varepsilon_e+\varepsilon_h \right).
\end{equation}
The quantity $\Sigma^{\sigma} \left( \omega \right)$ can be analytically calculated using a semiclassical expansion of the angular-momentum-restricted density of states \cite{Weick2005, Weick2006}, leading to
\begin{equation}
\label{eq0013}
 \gamma_{q\tau}^{\sigma, \mathrm{L}} = \frac{3 v_\mathrm{F}}{4 a} \left( \frac{\omega_0}{\omega_{q\tau}^{\sigma}} \right)^4 g\left( \frac{\hbar \omega_{q\tau}^{\sigma}}{E_\mathrm{F}} \right),
\end{equation}
where $v_\mathrm{F}$ and $E_\mathrm{F}$ are the Fermi velocity and energy of the metal, respectively. 
An explicit expression of the monotonically decaying function $g(z)\in\;]0,1]$ can be found in Eqs.\ (107) and (108) of Ref.\ \cite{Yannouleas1992} or in Eq.\ (B2) of Ref.~\cite{Weick2011}. Notably, the Landau damping decay rate is independent of the number of dimers $\mathcal{N}$ in the chain.

\begin{figure}[tb]
 \includegraphics[width=1.0\columnwidth]{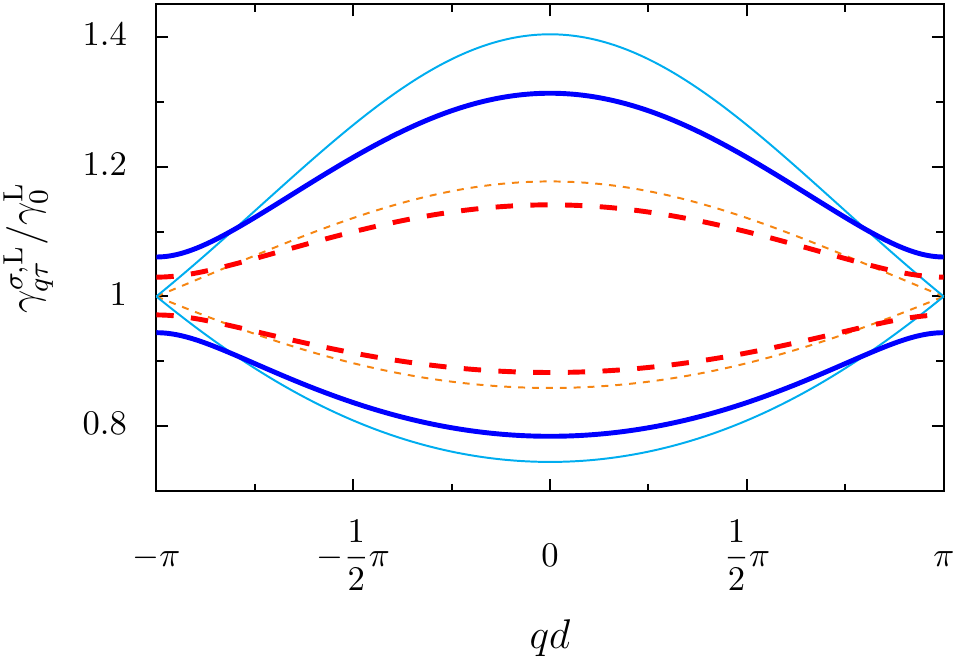}
 \caption{Landau damping decay rate from Eq.~\eqref{eq0013} as a function of momentum for the transverse (red and orange dashed lines, $ \sigma = x = y$) and longitudinal (blue and cyan solid lines, $ \sigma = z$) collective plasmonic modes for $\hbar \omega_0 / E_F = 0.5$. The thinner
(thicker)  lines correspond to $ d_1 = d_2 = 3 a $ ($ d_1 = 3.5 a $ and $ d_2 = 3 a $).}
 \label{fig6}
\end{figure}

\begin{figure*}[tb]
 \includegraphics[width=1.6\columnwidth]{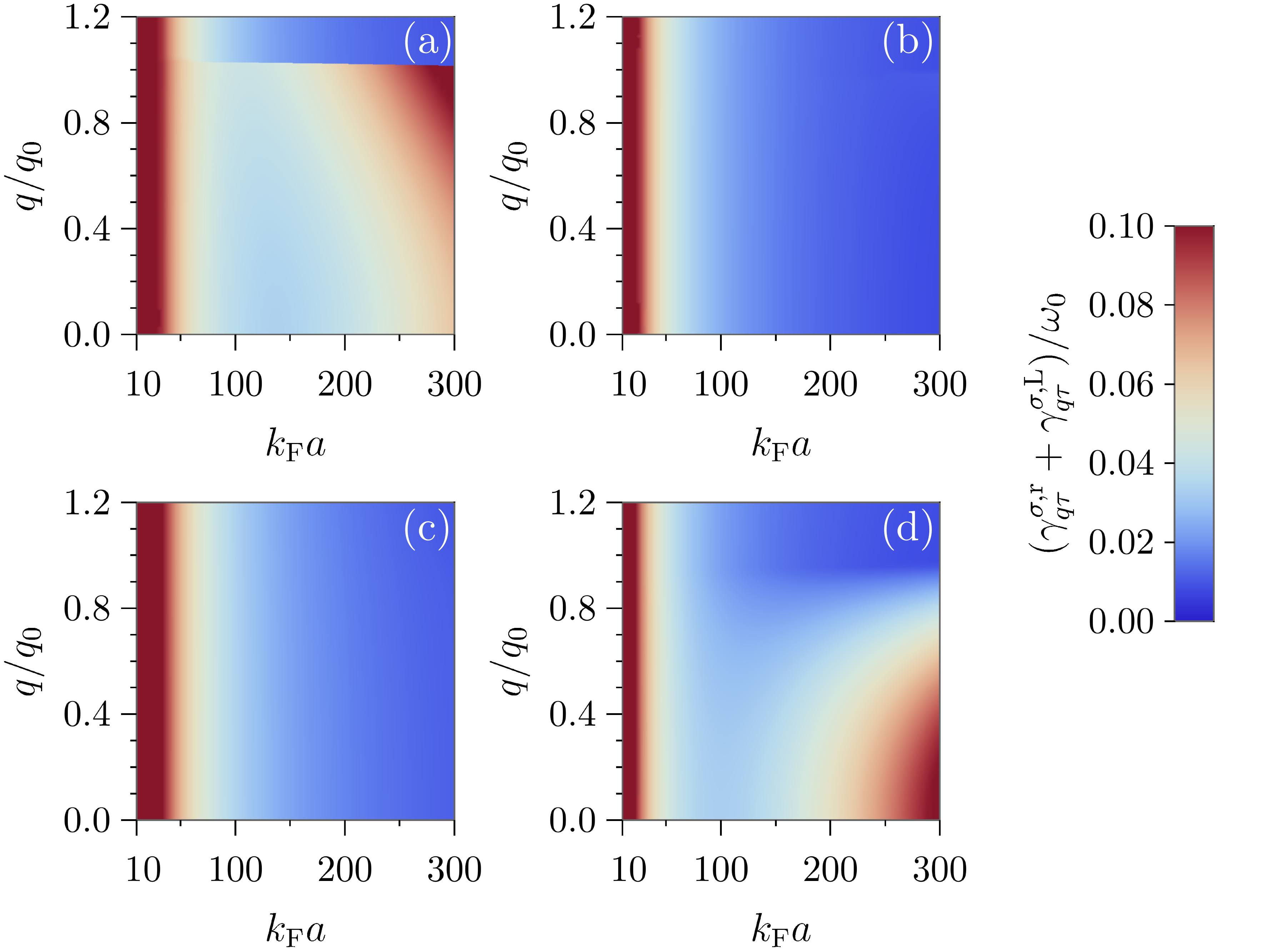}
 \caption{Sum of the radiative decay rate given by Eq.~\eqref{eq0008} and the Landau damping decay rate given by Eq.~\eqref{eq0013} as a function of both the scaled wavevector $q/q_0$ and the scaled nanoparticle radius $k_{\text{F}}a$, for a bipartite nanoparticle chain characterized by interparticle separations $d_1 = 3.5 a$ and $d_2 = 3a$. The parameters in the figure are $\hbar \omega_0 / E_{\mathrm{F}} = 0.47$ and $q_0/ k_{\mathrm{F}} = 1.1 \times 10^{-3}$, corresponding to a chain composed of Ag nanoparticles. The plasmonic bands under consideration are (a) the upper transverse band $\sigma =x, y$, $\tau = 1$, (b) the lower transverse band  $\sigma = x, y$, $\tau = -1$, (c) the upper longitudinal band  $\sigma = z$, $\tau = 1$, and (d) the lower longitudinal band  $\sigma = z$, $\tau = -1$. }
 \label{fig6b}
\end{figure*}

The Landau damping decay rate \eqref{eq0013} is plotted in Fig.\ \ref{fig6}, scaled with the single-nanoparticle result $\gamma_0^{\mathrm{L}} = 3 v_\mathrm{F} g( \hbar \omega_0 / E_\mathrm{F} ) / 4 a$ \cite{Kawabata1966, Yannouleas1992, Weick2005}. The results for the regular chain with $d_1 = d_2$ (thinner lines) and a typical dimerized chain with $d_1\neq d_2$ (thicker lines) for both the transverse (dashed red/orange lines) and longitudinal (solid blue/cyan lines) polarizations are shown. The Landau damping linewidths of the coupled plasmons are significantly different in magnitude depending on the band index $\tau$ due to the quartic dependence on the spectrum. While the $\tau=-1$ bands show decreasing Landau damping with increasing magnitude of the momentum, the opposite is true for the $\tau=+1$ bands. Therefore the center of the Brillouin zone displays the greatest disparity in decay rate between each of the two bands per polarization, being either appreciably higher or lower than the result of a single nanoparticle.
However, at the edge of the Brillouin zone, where the topologically-protected states described in Sec.\ \ref{sec5} reside at the midgap frequency $\omega_0$, the Landau damping decay rate is well approximated by the single-particle result $\gamma_0^{\mathrm{L}}$ 
[cf.\ Eq.\ \eqref{eq0013} with $\omega_{q\tau}^\sigma$ replaced by $\omega_0$].

\subsection{Experimental realization}

The preceding quantitative studies of both the radiative and Landau decay rates in Secs.~\ref{sec:radiative} and Sec.~\ref{sec:Landau}, respectively, allow us to investigate the optimal parameters of the bipartite chain of nanoparticles in order to maximize the lifetime of the collective plasmonic excitations. In Fig.\ \ref{fig6b}, we plot these aforementioned decay rates, scaled by the Mie frequency $\omega_0$, as a function of both $k_\mathrm{F}a$ ($k_\mathrm{F}$ denotes the Fermi wavevector) and the scaled momentum $q/q_0$. The parameters used in the figure correspond to a chain of silver nanoparticles with $\omega_{0} = 2.6 \text{eV} / \hbar$, $k_\mathrm{F}^{-1}=\unit[0.83]{\AA}$, $q_0^{-1}=\unit[76]{nm}$, and with interparticle separations $d_1 = 3.5 a$ and $d_2 = 3a$. The panels (a)-(d) in Fig.\ \ref{fig6b} display the competition between the radiative and Landau losses for each of the four plasmonic bands: (a) the upper transverse band ($\sigma=x,y$, $\tau=+1$), (b) the lower transverse band ($\sigma=x,y$, $\tau=-1$), (c) the upper longitudinal band ($\sigma=z$, $\tau=+1$), and (d) the lower longitudinal band ($\sigma=z$, $\tau=+1$). 
The drastic effects of radiative losses scaling with the nanoparticle size as $a^3$ [see Eq.\ \eqref{eq0008}] 
can be seen in the in-phase plasmonic bands, shown in panels (a) and (d), in stark contrast to the out-of-phase plasmonic bands, shown in panels (b) and (c). The two in-phase bands also exhibit the striking feature of a complete suppression of radiative losses outside the light cone $(q \gtrsim q_0)$, as was explicitly derived in Eq.\ \eqref{eq0008}. In particular, there is an abrupt drop in the total decay rate for 
the in-phase transverse modes at $q\simeq q_0$ [cf.\ Fig.\ \ref{fig6b}(a)] which is also observed in Fig.\ \ref{fig5} (dashed lines).
The common feature across all four types of plasmonic band is the quantum size effect of strong Landau damping for small nanoparticles, which graphically illustrates the result of Eq.\ \eqref{eq0013} and its distinctive $1/a$ scaling. It may be seen from Figs.~\ref{fig6b}(a) and \ref{fig6b}(d) that the optimal size of nanoparticle to probe bright modes of collective plasmonic excitations (of either polarization) is of the order of $k_\mathrm{F}a \simeq 120$, corresponding to $a\simeq\unit[10]{nm}$.

An estimate of the total plasmonic lifetime can be made by considering $1/\gamma_{q \tau}^{\sigma}$, where the total damping rate $\gamma_{q \tau}^{\sigma} =  \gamma_{q, \tau}^{\sigma, \mathrm{r}} +  \gamma_{q, \tau}^{\sigma, \mathrm{L}} + \gamma^{\mathrm{O}}$ includes the momentum-independent bulk Ohmic decay rate $\gamma^{\mathrm{O}}$. With a typical Ohmic damping rate $\gamma^{\mathrm{O}} = 0.027 \omega_0$ for silver nanoparticles \cite{charl89_ZPD}, the bright band plasmonic lifetime is approximately $\unit[4]{fs}$ for $a=\unit[10]{nm}$, such that experimental detection is a somewhat challenging task. However, an enhancement of the lifetime by a factor of two can be achieved by employing metallic nanoparticles with gain in order to compensate for Ohmic losses \cite{Maier2007}. Various methods of introducing gain into plasmonic systems exist, including embedding the nanoparticles in a dielectric environment with gain \cite{Lawandy2004, Noginov2007} or by cladding the nanoparticle core with a gain-material shell \cite{Noginov2009}. 

Regarding the topologically-protected edge states described in Sec.\ \ref{sec5}, a well-defined criterion for their experimental detection is that the gap 
$\Delta^\sigma$ formed at the edge of the Brillouin zone should typically be larger than the linewidth $\gamma^\mathrm{t}$ of the topological state.
As we discussed in Secs.\ \ref{sec:radiative} and \ref{sec:Landau}, the latter quantity can be approximated by $\gamma^\mathrm{t}\simeq\gamma^\mathrm{O}+\gamma_0^\mathrm{L}$, as it corresponds to a dark mode and therefore is immune to radiation damping. Dark modes can be excited via electron energy loss spectroscopy (EELS) \cite{Barrow2014, Barrow2016}, which enables the observation of plasmon modes which are not directly reachable optically. A simple estimate suggests that for a bipartite chain composed of Ag nanoparticles with radius $a=\unit[25]{nm}$ (for which Landau damping is negligible) and with $d_2=3a$ and $d_1\simeq4.6a$, the topological edge states corresponding to the longitudinal polarization should be well defined. The introduction of gain media would further increase the distinguishability of these edge states. Therefore, we conclude that the phenomena discussed in this work are within experimental accessibility, particularly for setups with optimally-sized metallic nanoparticles and appropriate gain media.

\section{Conclusion}
\label{sec7}

In this theoretical work, we have analyzed in detail the fundamental properties of collective plasmonic excitations in bipartite chains of spherical metallic nanoparticles. We have revealed that this system harbors an unusual form of collective excitations: topological bosonic Bogoliubov quasiparticles. Near the edge of the Brillouin zone, we have shown that these quasiparticles are described by a one-dimensional massive Dirac Hamiltonian. Therefore, we suggest bipartite chains of plasmonic nanoparticles as an ideal testbed to experimentally probe 1D Dirac-like bosons, which undergo exotic phenomena such as Klein tunneling. While our proposal complements the hosting of 2D Dirac-like bosons in honeycomb lattices \cite{Weick2013, Sturges2015}, it also has the notable advantage that the bipartite linear lattice may be easier to realize experimentally.

Having derived the full quantum eigenstates describing the bosonic Bogoliubov excitations, we were able to calculate the associated Zak phase of the system. This has allowed us to characterize two topologically distinct regimes, depending on the dimerization of the chain, and foreshadowed the presence of plasmonic edge states. We have studied the various types of topologically-protected edge states that may be formed and derived transparent formulas describing their behavior, which opens up a new avenue to manipulate light at the nanoscale.

Since combating radiative and nonradiative losses is a considerable challenge in plasmonics, we have undertaken a detailed study of the decay rates of plasmonic excitations in bipartite nanoparticle chains. We have provided analytic formulas describing losses in the system due to both radiation damping, which has uncovered the existence of both bright and dark plasmonic states, and Landau damping, a quantum size effect fundamental to the study of quantum metamaterials. We have therefore been able to estimate the lifetime of the collective plasmonic excitations in the system and hence the optimal parameters required for experimental detection. With the current trend of technological advances in the fabrication of increasing small metallic nanoparticle arrays with gain media, the topological and Dirac-like phenomena we have discussed should be within experimental reach.

\section*{Acknowledgments}

We are grateful to D.\ Weinmann for illuminating discussions and for 
his careful reading of the manuscript. 
We thank T.\ J.\ Sturges for technical support and T.\ Ojanen for useful correspondence.
We acknowledge financial support from the CNRS through the PICS program (Contract No.\ 6384 APAG) and from the ANR (Grant No.\ ANR-14-CE26-0005 Q-MetaMat).

\begin{appendix}
\section{\label{appendA}Collective plasmon dispersion: effect of dipole-dipole interactions beyond nearest neighbors}

In the plasmonic Hamiltonian \eqref{eq01}, we consider the interaction between nearest neighbors on the 1D bipartite lattice. However, as the near-field dipolar interaction decays as one over the cube of the interparticle distance, it is necessary to check the robustness of our results against the inclusion of interactions beyond nearest neighbors only. In this appendix, 
we show that the interaction between the nearest neighbors alone captures the relevant physics of the problem and that the plasmon dispersion~\eqref{eq08} is only slightly quantitatively modified by interactions beyond nearest neighbors.

In the following, we evaluate the collective plasmon dispersion classically by taking into account the interaction between an arbitrary number of neighbors on the bipartite lattice. Specifically, we consider the dipole-dipole interaction between a nanoparticle on site $(s, n)$ belonging to sublattice $s$ and dimer $n$ up to sites $(s, n\pm M)$ and $(\bar s, n\pm [M'-1])$ with $M,M'\geqslant1$ and where $\bar s=A$ $(B)$ for $s=B$ $(A)$. The plasmonic dispersion can thus be straightforwardly evaluated to yield
the general result
\begin{equation}
\label{eq123456}
 \omega_{q\tau}^{\sigma}(M, M') = \omega_0 \sqrt{1 + \xi_q^{\sigma}(M) + 2 \tau \frac{|\nu_q^{\sigma}(M')|}{\omega_0}}, 
\end{equation}
with 
\begin{equation}
\label{eq1234567}
 \xi_q^{\sigma}(M) = 2 \eta_{\sigma} \left(\frac{a}{d}\right)^3 \sum_{m=1}^{M} \frac{\cos(mqd)}{m^3}
 \end{equation}
and 
\begin{align}
\label{eq12345678}
 \nu_q^{\sigma}(M') =&\;\eta_{\sigma} \sum_{m=1}^{M'}\left\{
 \left[\frac{a}{(m-1)d+d_1}
 \right]^3
  \mathrm{e}^{ \mathrm{i} (m-1) q d} 
  \right.
  \nonumber\\
 &+\left.   \left[\frac{a}{(m-1)d+d_2}
 \right]^3 \mathrm{e}^{ - \mathrm{i} m q d}\right\}.
\end{align} 

\begin{figure}[tb]
 \includegraphics[width=1.0\columnwidth]{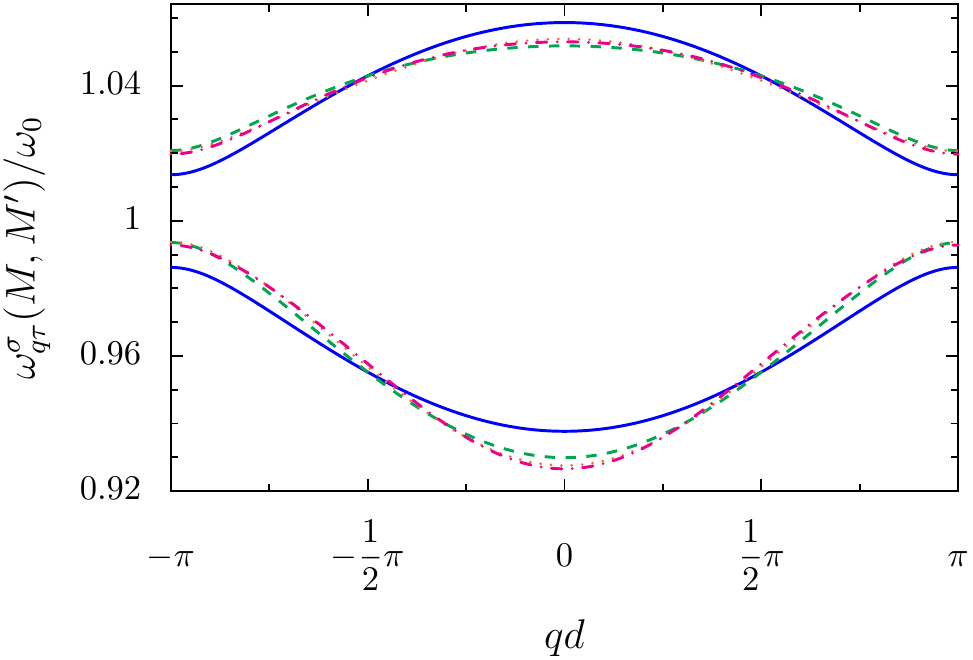}
 \caption{Collective plasmon dispersion relation with dipole-dipole interaction including nearest (solid blue lines), next nearest (dashed green lines), third nearest (dotted orange lines), and fourth nearest (dash-dotted magenta lines) neighbors, see Eq.\ \eqref{eq123456}. In the figure, $d_1 = 3 a$, $d_2 = 3.5 a$, and $\sigma = z$, corresponding to the longitudinal collective plasmon polarization.}
 \label{fig7}
\end{figure}

We plot the bandstructure \eqref{eq123456} in Fig.\ \ref{fig7} for different values of the number of neighbors taken into account (for clarity, we show only the longitudinal mode $\sigma=z$, since similar conclusions can be drawn from the two transverse modes
$\sigma=x,y$). The nearest neighbor case $\omega_{q\tau}^{\sigma}(0, 1)$ [corresponding to Eq.\ \eqref{eq08}] is shown by solid blue lines, while the cases with successively more nearest neighbors included, i.e., $\omega_{q\tau}^{\sigma}(1,1)$, $\omega_{q\tau}^{\sigma}(1,2)$, and $\omega_{q\tau}^{\sigma}(2,2)$ are represented by dashed green, dotted orange, and dash-dotted magenta lines, respectively. As can be seen from the figure, dipole-dipole interactions beyond the nearest neighbor contribution do not lead to significant qualitative changes in the collective plasmon dispersion relation. Quantitatively, when we discuss physics occurring close to the edge of the Brillouin zone we should renormalize the resonance frequency by a small amount $\omega_0 \to \omega_0 -\eta_\sigma(a/d)^3$. For simplicity, we thus limit ourselves to the discussion of dipole-dipole interaction effects between nearest neighbors in the main text.

\section{\label{appendB}Klein tunneling of massive Dirac-like collective plasmons}

In his appendix we consider the transmission problem of massive Dirac-like collective plasmons at the edge of the first Brillouin zone between two semi-infinite bipartite chains of metallic nanoparticles, denoted by L (for left) and R (for right), and described by two different resonance frequencies $\omega_0^{\mathrm{L, R}}$, respectively. Therefore, the four associated coupling constants arising from the dipole-dipole interaction
\begin{equation}
\label{eq:OmegaLR}
\Omega_{1, 2}^{\mathrm{L, R}}=\frac{\omega_0^{\mathrm{L, R}}}{2}\left(\frac{a}{d_{1,2}}\right)^3
\end{equation}
 are also (in general) different. For simplicity we here consider the case where 
 the interparticle spacings $d_1$ and $d_2$ are the same in the left and right chains. 
The Dirac-like collective plasmons at the edge of the first Brillouin zone $qd\simeq\pi$ are described by the spinor \eqref{eq13}, which pick up a label $j = (\mathrm{L, R})$ in each section of the bipartite chain, such that the quantity entering Eq.\ \eqref{eq13} becomes
\begin{equation}
\label{eq15a}
 \nu_{k_{j}}^{\sigma j} = \eta_{\sigma} \left( \Omega_1^j - \Omega_2^j + \mathrm{i}\, \Omega_2^j k_j d\right). 
\end{equation}

In our proposed setup, a notable contrast to the traditional Klein potential step scenario is that the change in on-site energy between the left and right parts of the chain $\omega_0^{\mathrm{L, R}}$  is intrinsically linked to a change in the mass term, as 
the coupling constants \eqref{eq:OmegaLR} are directly proportional to $\omega_0^{\mathrm{L, R}}$. Therefore, we effectively deal here with a combined `electrostatic-mass step', meaning in the Hamiltonian \eqref{eq14} both the customary diagonal terms $\propto \mathbbm{1}_2$ and the mass term $\propto \sigma_z$ change as the plasmonic excitation propagates from the left to the right section of the chain. 

The transmission probability $T$ can be calculated using the spinor wavefunctions describing incident, reflected, and transmitted plasmonic excitations, respectively. 
From conservation of energy, the planewaves' wavevectors are
\begin{equation}
\label{eq15b}
 k_j = \frac{\zeta_j}{ d \sqrt{\Omega_1^j \Omega_2^j}} \sqrt{\frac{\left(\omega - \omega_0^j\right)^2}{\eta_{\sigma}^2} - \left(\Omega_1^j-\Omega_2^j\right)^2}.
\end{equation}
Here $\omega$ is the frequency of the excitation propagating from left to right in a chain with $\omega_0^{\mathrm{R}} > \omega_0^{\mathrm{L}}$, and $\zeta_j = \pm 1$ arises to ensure conservation of the sign of the group velocity, which is not necessarily aligned with the wavevector. 
Applying the boundary conditions of continuity of both wavefunction components of the eigenspinors at the domain interface ($z=0$) and solving for the transmission coefficient, we arrive at the transmission probability
\begin{equation}
\label{eq15}
 T = \frac{\mathrm{Re} \left\{ \nu_{k_\mathrm{L}}^{\sigma\mathrm{L}} \left( \nu_{k_\mathrm{R}}^{\sigma\mathrm{R}} - {\nu_{k_\mathrm{R}}^{\sigma\mathrm{R}}}^* \right) \right\}  }
 { \mathrm{Re} \left\{\nu_{k_\mathrm{L}}^{\sigma\mathrm{L}} \nu_{k_\mathrm{R}}^{\sigma\mathrm{R}} \right\} - (\tau_{\mathrm{L}}/\tau_{\mathrm{R}}) \left|\nu_{k_\mathrm{L}}^{\sigma\mathrm{L}} \nu_{k_\mathrm{R}}^{\sigma\mathrm{R}}\right|  }.
\end{equation}

\begin{figure}[tb]
 \includegraphics[width=1.0\columnwidth]{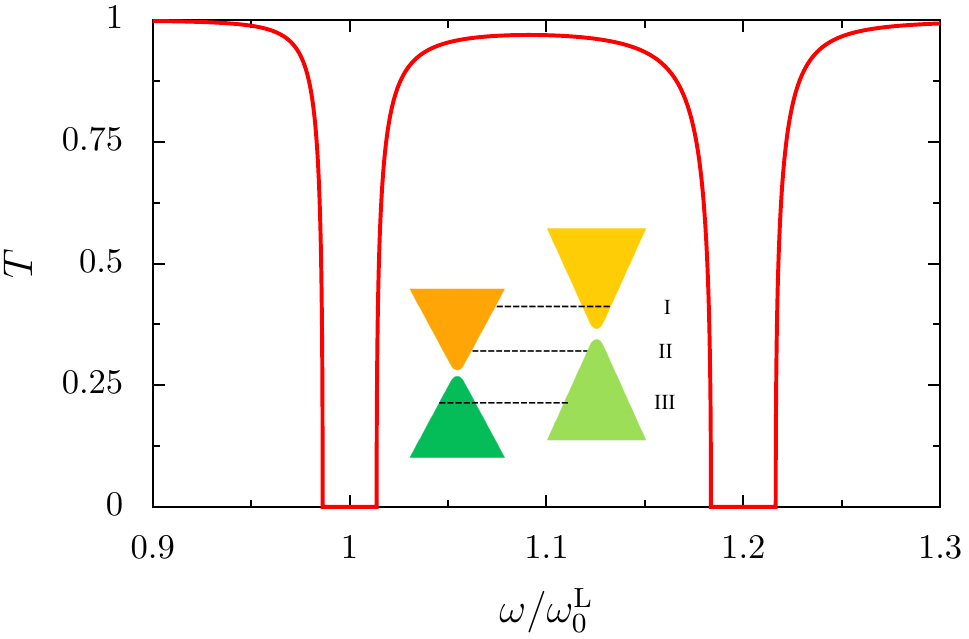}
 \caption{Transmission probability though a system with parameters $\omega_0^{\mathrm{R}} / \omega_0^{\mathrm{L}} = 1.2$, $ d_1 = 3 a $, $ d_2 = 3.5 a$, and $\sigma = z$. Inset: sketch of the spectrum of the quasiparticles. The excitations may be transmitted in three regimes (dotted lines) corresponding to: (I) electron-like to electron-like, (II) electron-like to hole-like, and (III) hole-like to hole-like states. }
 \label{fig8}
\end{figure}

 We plot in Fig.~\ref{fig8} the transmission probability of a longitudinal excitation ($\sigma=z$) as a function of excitation frequency in a chain with $\omega_0^{\mathrm{R}} > \omega_0^{\mathrm{L}}$. The problem can be partitioned into three different regimes of interest, as sketched in the inset of Fig.~\ref{fig8}. In region I (III), above $\omega_{\mathrm{R}} + |\eta_{\sigma}| |\Omega_1^{\mathrm{R}}-\Omega_2^{\mathrm{R}}|$ (below $\omega_{\mathrm{L}} - |\eta_{\sigma}| |\Omega_1^{\mathrm{L}}-\Omega_2^{\mathrm{L}}|$) the tunneling is from electron (hole)-like to electron (hole)-like states. However, in the interval II, where $\omega_{\mathrm{L}} + |\eta_{\sigma}| |\Omega_1^{\mathrm{L}}-\Omega_2^{\mathrm{L}}| < \omega < \omega_{\mathrm{R}} - |\eta_{\sigma}| |\Omega_1^{\mathrm{R}}-\Omega_2^{\mathrm{R}}|$, the effective quasiparticle species changes from electron-like to hole-like, thus relying on the Klein physics to transmit. Most notably, due to the small effective masses the transmission probabilities are nearly always approaching unity, while no transmission takes place in the gapped regions.

\section{\label{appendD}Edge states in bipartite chains with an odd number of nanoparticles}

In the main text, we consider bipartite chains consisting of an even number of nanoparticles and described by the Hamiltonian \eqref{eq01}. 
While bulk properties do not depend on this restriction (see Secs.\ \ref{sec3} and \ref{sec4}), the formation of edge states is 
highly influenced by the parity of the chain. 
For the sake of completeness, we derive  in this appendix expressions for the probability densities at each nanoparticle site corresponding to the midgap state at frequency $\omega_0$ for the case of a chain comprised of an odd number of nanoparticles.

\begin{figure}[b!]
 \includegraphics[width=1.0\columnwidth]{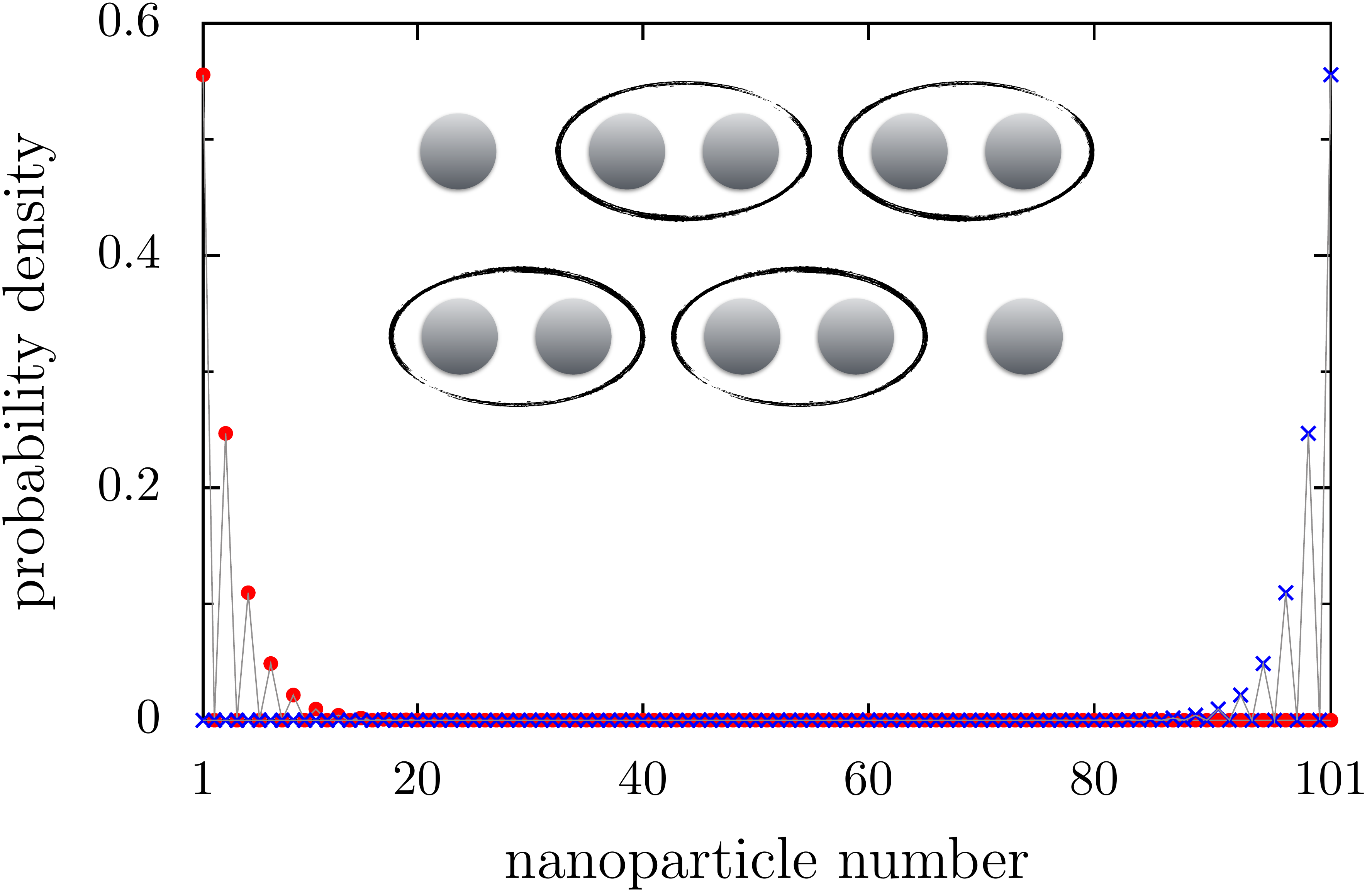}
 \caption{Plasmonic probability density corresponding to the midgap state at frequency $\omega_0$ at each site in a chain of 101 nanoparticles, for $\Omega_1 / \Omega_2 = 2/3$ (red dots) and $\Omega_1 / \Omega_2 = 3/2$ (blue crosses). The thin gray lines are guides for the eye. Inset: Sketches of a typical chain comprised of an odd number of nanoparticles, where the loops depict the favored dimerization for $\Omega_2 > \Omega_1$ (top) and $\Omega_1 > \Omega_2$ (bottom).}
 \label{fig9}
\end{figure}

Proceeding as in the case of an even number of nanoparticles (cf.\ Sec.\ \ref{sec:edge_states}) but adding an $A$ site on the right side of the chain, we now deal with a symmetric tridiagonal matrix $\sf{H}^{\sigma}_\mathrm{pl}$ of odd order, which is always guaranteed to support one eigenvalue at the midgap frequency $\omega_0$ \cite{Shin1997}. Here the boundary conditions arising at the edges of the chain are $B_1^\sigma = B^\sigma_{\mathcal{N}} = 0$, which together with the recurrence relation for $B$ sites imply the complete absence of a midgap plasmonic mode on one half of the nanoparticle sites. On the $A$ sites, nonvanishing probability densities arise as follows:
\begin{equation}
\label{eq8767678c}
\left|A_{n}^\sigma\right|^2 = \frac{(\Omega_1/\Omega_2)^2-1}{(\Omega_1/\Omega_2)^{2(\mathcal{N}+1)}-1}\left( \frac{\Omega_1}{\Omega_2} \right)^{2(n-1)}, \quad \Omega_1 < \Omega_2,
\end{equation}
\begin{equation}
\label{eq876765678c}
\left|A_{n}^\sigma\right|^2 =\frac{(\Omega_2/\Omega_1)^2-1}{(\Omega_2/\Omega_1)^{2(\mathcal{N}+1)}-1}\left( \frac{\Omega_2}{\Omega_1} \right)^{2(\mathcal{N}+1-n)}, \quad \Omega_1 > \Omega_2, 
\end{equation}
with $n \in [1, \mathcal{N}+1]$. 
Thus, highly localized collective plasmons occur at one edge only: either the left [Eq.~\eqref{eq8767678c}] or right [Eq.~\eqref{eq876765678c}] end of the chain, determined by the critical ratio $\Omega_1/\Omega_2$, as exemplified in Fig.\ \ref{fig9}. Such an outcome can be naturally explained with the aid of the sketches in the inset of the figure. 
Indeed, one nanoparticle is always left over when the chain dimerizes. Depending on the coupling ratio $\Omega_1/\Omega_2$ 
one can choose the remaining unpaired nanoparticle, and hence the edge state, to be at either side of the chain.

\end{appendix}



\begin{thebibliography}{100}

\bibitem{Veselago1968}V.\ G.\ Veselago, The electrodynamics of substances with simultaneously negative values of $\epsilon$ and $\mu$, \href{http://dx.doi.org/10.1070/PU1968v010n04ABEH003699}{Sov. Phys. Usp. \textbf{10}, 509 (1968)}.

\bibitem{Schurig2006}D.\ Schurig, J.\ J.\ Mock, B.\ J.\ Justice, S.\ A.\ Cummer, J.\ B.\ Pendry, A.\ F.\ Starr, and D.\ R.\ Smith, Metamaterial electromagnetic cloak at microwave frequencies, \href{http://dx.doi.org/10.1126/science.1133628}{Science \textbf{314}, 977 (2006)}.

\bibitem{Tsakmakidis2007}K.\ L.\ Tsakmakidis, A.\ D.\ Boardman, and O.\ Hess, `Trapped rainbow' storage of light in metamaterials, \href{http://dx.doi.org/10.1038/nature06285}{Nature \textbf{450}, 397 (2007)}.

\bibitem{Zhang2008}X.\ Zhang and Z.\ Liu, Superlenses to overcome the diffraction limit, \href{http://dx.doi.org/10.1038/nmat2141}{Nat. Mater. \textbf{7}, 435 (2008)}. 

\bibitem{Bertsch1994}G.\ F.\ Bertsch and R.\ A.\ Broglia, \textit{Oscillations in Finite Quantum Systems} (Cambridge University Press, Cambridge, 1994).

\bibitem{Kreibig1995}U.\ Kreibig and M.\ Vollmer, \textit{Optical Properties of Metal Clusters} (Springer-Verlag, Berlin, 1995).

\bibitem{Maier2007}S.\ A.\ Maier, \textit{Plasmonics: fundamentals and applications} (Springer-Verlag, Berlin, 2007).

\bibitem{Barnes2003}W.\ L.\ Barnes, A.\ Dereux, and T.\ W.\ Ebbesen, Surface plasmon subwavelength optics, \href{http://dx.doi.org/10.1038/nature01937}{Nature \textbf{424}, 824 (2003)}.

\bibitem{Bergman2003}D.\ J.\ Bergman and M.\ I.\ Stockman, Surface plasmon amplification by stimulated emission of radiation: quantum generation of coherent surface plasmons in nanosystems, \href{http://dx.doi.org/10.1103/PhysRevLett.90.027402}{Phys. Rev. Lett. \textbf{90}, 027402 (2003)}.

\bibitem{Pacifici2007}D.~Pacifici, H.~J.~Lezec, and H.~A.~Atwater, All-optical modulation by plasmonic excitation of CdSe quantum dots, \href{http://dx.doi.org/10.1038/nphoton.2007.95}{Nat. Photon. \textbf{1}, 402 (2007)}.

\bibitem{Anker2008}J.\ N.\ Anker, W.\ P.\ Hall, O.\ Lyandres, N.\ C.\ Shah, J.\ Zhao, and R.\ P.\ Van Duyne, Biosensing with plasmonic nanosensors, \href{http://dx.doi.org/10.1038/nmat2162}{Nat. Mater. \textbf{7}, 442 (2008)}.

\bibitem{Pala2008}R.\ A.\ Pala, K.\ T.\ Shimizu, N.\ A.\ Melosh, and M.\ L.\ Brongersma, A nonvolatile plasmonic switch employing photochromic molecules, \href{http://dx.doi.org/10.1021/nl0808839}{Nano Lett. \textbf{8}, 1506 (2008)}.

\bibitem{Zijlstra2015}P.~Zijlstra, J.~W.~M.~Chon, and M.~Gu, Five-dimensional optical recording mediated by surface plasmons in gold nanorods, \href{http://dx.doi.org/10.1038/nature08053}{Nature \textbf{459}, 410 (2009)}.

\bibitem{Stockman2004}M.~I.~Stockman, Nanofocusing of optical energy in tapered plasmonic waveguides, \href{http://dx.doi.org/10.1103/PhysRevLett.93.137404}{Phys. Rev. Lett. \textbf{93}, 137404 (2004)}.

\bibitem{Oulton2008}R.\ F.\ Oulton, V.\ J.\ Sorger, D.\ A.\ Genov, D.\ F.\ P.\ Pile, and X.\ Zhang, A hybrid plasmonic waveguide for subwavelength confinement and long-range propagation, \href{http://dx.doi.org/10.1038/nphoton.2008.131}{Nat. Photon. \textbf{2}, 496 (2008)}. 

\bibitem{Quinten1998}M.~Quinten, A.~Leitner, J.~R.~Krenn, and F.~R.~Aussenegg, Electromagnetic energy transport via linear chains of silver nanoparticles, \href{http://dx.doi.org/10.1364/OL.23.001331}{Opt. Lett. \textbf{23}, 1331 (1998)}.

\bibitem{Meinzer2014}N.~Meinzer, W.~L.~Barnes, and I.~R.~Hooper, Plasmonic meta-atoms and metasurfaces, \href{http://dx.doi.org/10.1038/nphoton.2014.247}{Nat. Photon. \textbf{8}, 889 (2014)}. 

\bibitem{Brongersma2000}M.~L.~Brongersma, J.~W.~Hartman, and H.~A.~Atwater, Electromagnetic energy transfer and switching in nanoparticle chain arrays below the diffraction limit, \href{http://dx.doi.org/10.1103/PhysRevB.62.R16356}{Phys. Rev. B \textbf{62}, R16356 (2000).}

\bibitem{Maier2003b}S.~A.~Maier, P.~G.~Kik, and H.~A.~Atwater, Optical pulse propagation in metal nanoparticle chain waveguides, \href{http://dx.doi.org/10.1103/PhysRevB.67.205402}{Phys. Rev. B \textbf{67}, 205402 (2003)}.

\bibitem{Weber2004}W.~H.~Weber and G.~W.~Ford, Propagation of optical excitations by dipolar interactions in metal nanoparticle chains, \href{http://dx.doi.org/10.1103/PhysRevB.70.125429}{Phys. Rev. B \textbf{70}, 125429 (2004)}.

\bibitem{Park2004}S.~Y.~Park and D.~Stroud, Surface-plasmon dispersion relations in chains of metallic nanoparticles: an exact quasistatic calculation, \href{http://dx.doi.org/10.1103/PhysRevB.69.125418}{Phys. Rev. B \textbf{69}, 125418 (2004)}.

\bibitem{Citrin2004}D.~S.~Citrin, Coherent excitation transport in metal-nanoparticle chains, \href{http://dx.doi.org/10.1021/nl049679l}{Nano Lett. \textbf{4}, 1561 (2004)}.

\bibitem{Markel2007}V.~A.~Markel and A.~K.~Sarychev, Propagation of surface plasmons in ordered and disordered chains of metal nanospheres, \href{http://dx.doi.org/10.1103/PhysRevB.75.085426}{Phys. Rev. B \textbf{75}, 085426 (2007)}.

\bibitem{Lee2012}C.~Lee, M.~Tame, J.~Lim, and J.~Lee, Quantum plasmonics with a metal nanoparticle array, \href{http://dx.doi.org/10.1103/PhysRevA.85.063823}{Phys. Rev. A \textbf{85}, 063823 (2012)}.

\bibitem{Pino2014}J.~del~Pino, J.~Feist, F.~J.~Garcia-Vidal, and J.~J.~Garcia-Ripoll, Entanglement detection in coupled particle plasmons, \href{http://dx.doi.org/10.1103/PhysRevLett.112.216805}{Phys. Rev. Lett. \textbf{112}, 216805 (2014)}.

\bibitem{Brandstetter2016}A.\ Brandstetter-Kunc, G.\ Weick, C.\ A.\ Downing, D.\ Weinmann, and R.\ A.\ Jalabert, Nonradiative limitations to plasmon propagation in chains of metallic nanoparticles, 
\href{http://dx.doi.org/10.1103/PhysRevB.94.205432}{Phys. Rev. B \textbf{94}, 205432 (2016)}.

\bibitem{Krenn1999}J.\ R.\ Krenn, A.\ Dereux, J.\ C.\ Weeber, E.\ Bourillot, Y.\ Lacroute, J.\ P.\ Goudonnet, G.\ Schider, W.\ Gotschy, A.\ Leitner, F.\ R.\ Aussenegg, and C.\ Girard, Squeezing the optical near-field zone by plasmon coupling of metallic nanoparticles, \href{http://dx.doi.org/10.1103/PhysRevLett.82.2590}{Phys. Rev. Lett. \textbf{82}, 2590 (1999)}.

\bibitem{Maier2002}S.\ A.\ Maier, M.\ L.\ Brongersma, P.\ G.\ Kik, and H.\ A.\ Atwater, Observation of near-field coupling in metal nanoparticle chains using far-field polarization spectroscopy, \href{http://dx.doi.org/10.1103/PhysRevB.65.193408}{Phys. Rev. B \textbf{65}, 193408 (2002)}.

\bibitem{Maier2003a}S.~A.~Maier, P.~G.~Kik, H.~A.~Atwater, S.~Meltzer, E.~Harel, B.~E.~Koel, and A.~A.~G.~Requicha, Local detection of electromagnetic energy transport below the diffraction limit in metal nanoparticle plasmon waveguides, \href{http://dx.doi.org/10.1038/nmat852}{Nat. Mater. \textbf{2}, 229 (2003)}.

\bibitem{Solis2012}D.~Solis Jr., B.~Willingham, S.~L.~Nauert, L.~S.~Slaughter, J.~Olson, P.~Swanglap, A.~Paul, W-S.~Chang, and S.~Link, Electromagnetic energy transport in nanoparticle chains via dark plasmon modes, \href{http://dx.doi.org/10.1021/nl2039327}{Nano Lett. \textbf{12}, 1349 (2012)}.

\bibitem{Apuzzo2013}A.~Apuzzo, M.~Fevrier, R.~Salas-Montiel, A.~Bruyant, A.~Chelnokov, G.~Lerondel, B.~Dagens, and S.~Blaize, Observation of near-field dipolar interactions involved in a metal nanoparticle chain waveguide, \href{http://dx.doi.org/10.1021/nl304164y}{Nano Lett. \textbf{13}, 1000 (2013)}.

\bibitem{Barrow2014}S.~J.~Barrow, D.~Rossouw, A.~M.~Funston, G.~A.~Botton, and P.~Mulvaney, Mapping bright and dark modes in gold nanoparticle chains using electron energy loss spectroscopy, \href{http://dx.doi.org/10.1021/nl5009053}{Nano Lett. \textbf{14}, 3799 (2014)}.

\bibitem{poddu14_ACSPhoton}
A. Poddubny, A. Miroshnichenko, A. Slobozhanyuk, and Y. Kivshar, 
Topological Majorana states in zigzag chains of plasmonic nanoparticles, 
\href{http://dx.doi.org/0.1021/ph4000949}
{ACS Photonics \textbf{1}, 101 (2014)}.

\bibitem{Ling2015}C.~W.~Ling, M.~Xiao, C.~T.~Chan, S.~F.~Yu, and K.~H.~Fung, 
Topological edge plasmon modes between diatomic chains of plasmonic nanoparticles, \href{http://dx.doi.org/10.1364/OE.23.002021}{Opt. Express \textbf{23}, 2021 (2015)}.

\bibitem{Kanshu2012}
A. Kanshu, C. E. R\"uter, D. Kip, V. Shandarov, P. P. Beli\v{c}ev, I. Ili\'c, and M.\ Stepi\'c, 
Observation of discrete gap solitons in one-dimensional waveguide arrays with alternating
spacings and saturable defocusing nonlinearity, 
\href{http://dx.doi.org/10.1364/OL.37.001253}
{Opt. Lett. \textbf{37}, 1253 (2012)}.

\bibitem{Schomerus2013}
H. Schomerus, 
Topologically protected midgap states in complex photonic lattices, 
\href{http://dx.doi.org/10.1364/OL.38.001912}
{Opt. Lett. \textbf{38}, 1912 (2013)}.

\bibitem{Poli2015}
C. Poli, M. Bellec, U. Kuhl, F. Mortessagne, and H. Schomerus, 
Selective enhancement of topologically induced interface states in a dielectric resonator chain, 
\href{http://dx.doi.org/10.1038/ncomms7710}
{Nat. Commun. \textbf{6}, 6710 (2015)}.

\bibitem{Slobo2015}
A. P. Slobozhanyuk, A. N. Poddubny, A. E. Miroshnichenko, P. A. Belov, and Y. S. Kivshar, 
Subwavelength topological edge states in optically resonant dielectric structures, 
\href{http://dx.doi.org/10.1103/PhysRevLett.114.123901}
{Phys. Rev. Lett. \textbf{114}, 123901 (2015)}.

\bibitem{Blanco2016}
A. Blanco-Redondo, I. Andonegui, M. J. Collins, G. Harari, Y. Lumer, M. C. Rechtsman, B. I. Eggleton, and M. Segev, 
Topological optical waveguiding in silicon and the transition between topological and trivial defect states, 
\href{http://dx.doi.org/10.1103/PhysRevLett.116.163901}
{Phys. Rev. Lett. \textbf{116}, 163901 (2016)}; 
\href{http://dx.doi.org/10.1103/PhysRevLett.117.129901}
{\textbf{117}, 129901(E) (2016)}.

\bibitem{Solny2016}
D. D. Solnyshkov, A. V. Nalitov, and G. Malpuech, 
Kibble-Zurek mechanism in topologically nontrivial zigzag chains of polariton micropillars, 
\href{http://dx.doi.org/10.1103/PhysRevLett.116.046402}
{Phys. Rev. Lett. \textbf{116}, 046402 (2016)}.

\bibitem{Solny2016b}
D. D. Solnyshkov, O. Bleu, B. Teklu, and G. Malpuech, 
Chirality of topological gap solitons in bosonic dimer chains, 
\href{http://dx.doi.org/10.1103/PhysRevLett.118.023901}
{Phys. Rev. Lett. \textbf{118}, 023901 (2017)}.

\bibitem{liu16_preprint}
C. Liu, M.V. Gurudev Dutt, and D. Pekker, 
Robust manipulation of light using topologically protected plasmonic modes, 
\href{http://arxiv.org/abs/1610.08941}
{arXiv:1610.08941}.

\bibitem{Lu2014}
L. Lu, J. D. Joannopoulos, and M. Solja\v{c}i\'c, 
Topological photonics, 
\href{http://dx.doi.org/10.1038/NPHOTON.2014.248}
{Nature Photon. \textbf{8}, 821 (2014)}.

\bibitem{hasan10_RMP}
M. Z. Hasan and C. L. Kane, Colloquium: Topological insulators, 
\href{http://dx.doi.org/10.1103/RevModPhys.82.3045}{Rev. Mod. Phys. \textbf{82}, 3045 (2010)}.

\bibitem{Tame2013}
M. S. Tame, K. R. McEnery, \c{S.} K. \"Ozdemir, J. Lee, S. A. Maier, and M. S. Kim, 
Quantum plasmonics, 
\href{http://dx.doi.org/10.1038/NPHYS2615}
{Nature Phys. \textbf{9}, 329 (2013)}.

\bibitem{Klein1929}O.~Klein, Die reflexion von elektronen an einem potentialsprung nach der relativistischen dynamik von Dirac, 
\href{http://dx.doi.org/10.1007/BF01339716}{Z. Phys. \textbf{53}, 157 (1929)}.

\bibitem{katsn06_NatPhys}
M. I. Katsnelson, K. S. Novoselov, and A. K. Geim, 
Chiral tunneling and the Klein paradox in graphene, 
\href{http://dx.doi.org/10.1038/nphys384}
{Nat. Phys. \textbf{2}, 620 (2006)}.

\bibitem{Qi2008}X.-L.~Qi, T.~L.~Hughes, and S.-C.~Zhang, Topological field theory of time-reversal invariant insulators, \href{http://dx.doi.org/10.1103/PhysRevB.78.195424}{Phys. Rev. B \textbf{78}, 195424 (2008)}.

\bibitem{Jackiw1976}R. Jackiw and C. Rebbi, Solitons with fermion number $1/2$, \href{http://dx.doi.org/10.1103/PhysRevD.13.3398}{Phys. Rev. D \textbf{13}, 3398 (1976)}.

\bibitem{Khurgin2015}J.~B.~Khurgin, How to deal with the loss in plasmonics and metamaterials, \href{http://dx.doi.org/10.1038/nnano.2014.310}{Nat. Nanotechnol. \textbf{10}, 2 (2015)}.

\bibitem{Kawabata1966}A.~Kawabata and R.~Kubo, Electronic properties of fine metallic particles. II. Plasma resonance absorption, \href{http://dx.doi.org/10.1143/JPSJ.21.1765}{J. Phys. Soc. Jpn. \textbf{21}, 1765 (1966)}. 

\bibitem{Weick2013}G.~Weick, C.~Woollacott, W.~L.~Barnes, O.~Hess, and E.~Mariani, Dirac-like plasmons in honeycomb lattices of metallic nanoparticles, \href{http://dx.doi.org/10.1103/PhysRevLett.110.106801}{Phys. Rev. Lett. \textbf{110}, 106801 (2013)}.

\bibitem{Sturges2015}T.~J.~Sturges, C.~Woollacott, G.~Weick and E.~Mariani, Dirac plasmons in bipartite lattices of metallic nanoparticles, \href{http://dx.doi.org/10.1088/2053-1583/2/1/014008}{2D Mater. \textbf{2}, 014008 (2015)}.

\bibitem{Weick2015}G.~Weick and E.~Mariani, Tunable plasmon polaritons in arrays of interacting metallic nanoparticles, \href{http://dx.doi.org/10.1140/epjb/e2014-50658-2}{Eur. Phys. J. B \textbf{88}, 7 (2015)}.

\bibitem{Lamowski2016}S.~Lamowski, F.~Hellbach, E.~Mariani, G.~Weick, and F.~Pauly, Plasmon polaritons in cubic lattices of spherical metallic nanoparticles, \href{https://arxiv.org/abs/1606.04897}{arXiv:1606.04897}.

\bibitem{cohen}
C. Cohen-Tannoudji, J. Dupont-Roc, and G. Grynberg, 
\textit{Atom-Photon Interactions: Basic Processes and Applications} (Wiley- VCH, New York, 1992).

\bibitem{craig}
D. P. Craig and T. Thirunamachandran, 
\textit{Molecular Quantum Electrodynamics} 
(Academic Press, 1984, London).

\bibitem{Su1979}W.~P.~Su, J.~R.~Schrieffer, and A.~J.~Heeger, Solitons in polyacetylene, \href{http://dx.doi.org/10.1103/PhysRevLett.42.1698}{Phys. Rev. Lett. \textbf{42}, 1698 (1979)}.

\bibitem{Su1980}W.~P.~Su, J.~R.~Schrieffer, and A.~J.~Heeger, Soliton excitations in polyacetylene, \href{http://dx.doi.org/10.1103/PhysRevB.22.2099}{Phys. Rev. B \textbf{22}, 2099 (1980)}.

\bibitem{Yannouleas1992}C.~Yannouleas and R.~A.~Broglia, Landau damping and wall dissipation in large metal clusters, \href{http://dx.doi.org/10.1016/0003-4916(92)90340-R}{Ann. Phys. \textbf{217}, 1 (1992)}.

\bibitem{Weick2005}G.~Weick, R.~A.~Molina, D.~Weinmann, and R.~A.~Jalabert, Lifetime of the first and second collective excitations in metallic nanoparticles, \href{http://dx.doi.org/10.1103/PhysRevB.72.115410}{Phys. Rev. B \textbf{72}, 115410 (2005)}.

\bibitem{Brandstetter2015}A.~Brandstetter-Kunc, G.~Weick, D.~Weinmann, and R.~A.~Jalabert, Decay of dark and bright plasmonic modes in a metallic nanoparticle dimer, \href{http://dx.doi.org/10.1103/PhysRevB.91.035431}{Phys. Rev. B \textbf{91}, 035431 (2015)}; \href{http://dx.doi.org/10.1103/PhysRevB.92.199906}{\textbf{92}, 199906(E) (2015)}.

\bibitem{footnote1} 
It is known, e.g., from studies of crystalline structure with more than one atom per unit cell 
[see C.\ Bena and G.\ Montambaux, Remarks on the tight-binding model of graphene, \href{http://dx.doi.org/10.1088/1367-2630/11/9/095003}{New J. Phys. \textbf{11}, 095003 (2009)}], 
that while the expectation values of operators describing physical quantities is independent of the basis chosen, the form of the operators may depend on the basis. Here we made this choice of transformation to ensure that we arrive at the most transparent form of both the Zak phase, which is dependent on the choice of the unit cell, and also the effective Dirac Hamiltonian which both arise later in this work.

\bibitem{footnote2}The very form of the Bogoliubov transformation \eqref{eq05} appearing in the Hamiltonian~\eqref{eq07} exposes the unusual nature of the excitations as a nontrivial mixture of both creation and annihilation operators of the original plasmonic particles. Indeed, the vacuum of the Bogoliubov excitations $\beta_{q\tau}^{\sigma} \ket{\mathrm{vac}} = 0$ is not the vacuum of the original $(a_q^{\sigma}, b_q^{\sigma})$ plasmonic excitations. Instead, $\ket{\mathrm{vac}}$ involves pairs of $(a_q^{\sigma}, b_q^{\sigma})$ excitations which are virtually excited by the dipolar interaction. 

\bibitem{Peierls1955}R.~E.~Peierls, \textit{Quantum Theory of Solids} (Oxford University Press, Oxford, 1955).

\bibitem{Jackiw1983}R.~Jackiw and G.~Semenoff, Continuum quantum field theory for a linearly conjugated diatomic polymer with fermion fractionization, \href{http://dx.doi.org/10.1103/PhysRevLett.50.439}{Phys. Rev. Lett. \textbf{50}, 439 (1983)}.

\bibitem{Kane1997}C.~L.~Kane and E.~J.~Mele, Size, shape, and low energy electronic structure of carbon nanotubes, \href{http://dx.doi.org/10.1103/PhysRevLett.78.1932}{Phys. Rev. Lett. \textbf{78}, 1932 (1997)}.

\bibitem{Neto2009}
A. H. Castro Neto, F. Guinea, N. M. R. Peres, K. S. Novoselov, and A. K. Geim, 
The electronic properties of graphene, 
\href{http://dx.doi.org/10.1103/RevModPhys.81.109}
{Rev. Mod. Phys. \textbf{81}, 109 (2009).}

\bibitem{Banerjee2016}S.~Banerjee, J.~Fransson, A.~M.~Black-Schaffer, H.~Agren, and A.~V.~Balatsky, Granular superconductor in a honeycomb lattice as a realization of bosonic Dirac material, \href{http://dx.doi.org/10.1103/PhysRevB.93.134502}{Phys. Rev. B \textbf{93}, 134502 (2016)}.

\bibitem{Brack1993}M.~Brack, The physics of simple metal clusters: self-consistent jellium model and semiclassical approaches, \href{http://dx.doi.org/10.1103/RevModPhys.65.677}{Rev. Mod. Phys. \textbf{65}, 677 (1993)}.

\bibitem{Weick2006}G.~Weick, G.-L.~Ingold, R.~A.~Jalabert, and D.~Weinmann, Surface plasmon in metallic nanoparticles: renormalization effects due to electron-hole excitations, \href{http://dx.doi.org/10.1103/PhysRevB.74.165421}{Phys. Rev. B \textbf{74}, 165421 (2006)}.

\bibitem{Berry1984}M.~V.~Berry, Quantal phase factors accompanying adiabatic changes, \href{http://dx.doi.org/10.1098/rspa.1984.0023}{Proc. R. Soc. London Ser. A \textbf{392}, 45 (1984)}.

\bibitem{Zak1989}J.~Zak, Berry's phase for energy bands in solids, \href{http://dx.doi.org/10.1103/PhysRevLett.62.2747}{Phys. Rev. Lett. \textbf{62}, 2747 (1989)}.

\bibitem{Delplace2011}P.~Delplace, D.~Ullmo, and G.~Montambaux, Zak phase and the existence of edge states in graphene, \href{http://dx.doi.org/10.1103/PhysRevB.84.195452}{Phys. Rev. B \textbf{84}, 195452 (2011)}.

\bibitem{footnote3}
We note that these bosonic eigenstates are normalized according to an inner product with a metric, $\expect{\psi_{q\tau}^{\sigma}|\psi_{q\tau}^{\sigma}}_{\mathcal{J}} = \expect{\psi_{q\tau}^{\sigma}| \mathcal{J} |\psi_{q\tau}^{\sigma}} = 1$, where $\mathcal{J} = \sigma_z \otimes \mathbbm{1}_2$.  This prescription arises due to the necessity of observing bosonic statistics, see Refs.\ \cite{Tsallis1978, Colpa1978, Kawaguchi2012}.

\bibitem{Tsallis1978}C.~Tsallis, Diagonalization methods for the general bilinear Hamiltonian of an assembly of bosons, \href{http://dx.doi.org/10.1063/1.523549}{J. Math. Phys. \textbf{19}, 277 (1978)}.

\bibitem{Colpa1978}J.~H.~P.~Colpa, Diagonalization of the quadratic boson Hamiltonian, \href{http://dx.doi.org/10.1016/0378-4371(78)90160-7}{Physica A \textbf{93}, 327 (1978)}.

\bibitem{Kawaguchi2012}Y.~Kawaguchi and M.~Ueda, Spinor Bose-Einstein condensates, \href{http://dx.doi.org/10.1016/j.physrep.2012.07.005}{Phys. Rep. \textbf{520}, 253 (2012)}.

\bibitem{footnote5}A method to diagonalize a general bosonic quadratic Hamiltonian, which utilizes paraunitary matrices in order to conform to the bosonic commutation relations, is described in Refs.\ \cite{Tsallis1978, Colpa1978, Kawaguchi2012}. The required paraunitary matrices are obtained by a Cholesky decomposition of the bosonic Hamiltonian written in Bogoliubov-de Gennes form.

\bibitem{footnote:length}
We obtain already a good agreement between our numerical calculations and the analytical result \eqref{eq88} for chains consisting of $10$ nanoparticles. 

\bibitem{Asboth}
J. K. Asb\'oth, L. Oroszl\'any, and A. P\'alyi, 
\textit{A Short Course on Topological Insulators}
(Springer, Berlin, 2016).

\bibitem{Weick2011}G.~Weick and D.~Weinmann, Lifetime of the surface magnetoplasmons in metallic nanoparticles, \href{http://dx.doi.org/10.1103/PhysRevB.83.125405}{Phys. Rev. B \textbf{83}, 125405 (2011)}.

\bibitem{charl89_ZPD}
K.-P. Charl\'e, W. Schulze, and B. Winter, 
The size dependent shift of the surface plasmon absorption band of small spherical metal particles, 
\href{http://dx.doi.org/10.1007/BF01427000}
{Z.Phys. D \textbf{12}, 471 (1989)}.

\bibitem{Lawandy2004}N.~M.~Lawandy, Localized surface plasmon singularities in amplifying media, \href{http://dx.doi.org/10.1063/1.1825058}{Appl. Phys. Lett. \textbf{85}, 5040 (2004)}.

\bibitem{Noginov2007}M.~A.~Noginov, G.~Zhu, M.~Bahoura, J.~Adegoke, C.~Small, B.~A.~Ritzo, V.~P.~Drachev, and V.~M.~Shalaev, The effect of gain and absorption on surface plasmons in metal nanoparticles, \href{http://dx.doi.org/10.1007/s00340-006-2401-0}{Appl. Phys. B \textbf{86}, 455 (2007)}.

\bibitem{Noginov2009}M.~A.~Noginov, G.~Zhu, A.~M.~Belgrave, R.~Bakker, V.~M.~Shalaev, E.~E.~Narimanov, S.~Stout, E.~Herz, T.~Suteewong, and U.~Wiesner, Demonstration of a spaser-based nanolaser, \href{http://dx.doi.org/10.1038/nature08318}{Nature \textbf{460}, 1110 (2009) }.

\bibitem{Barrow2016}
S.\ J.\ Barrow, S.\ M.\ Collins, D.\ Rossouw, A.\ M.\ Funston, G.\ A.\ Botton, P.\ A.\ Midgley, and P.\ Mulvaney, 
Electron energy loss spectroscopy investigation into symmetry in gold trimer and tetramer plasmonic nanoparticle structures,  
\href{http://dx.doi.org/10.1021/acsnano.6b03796}{ACS Nano \textbf{10}, 8552 (2016)}.

\bibitem{Shin1997}B.~C.~Shin, A formula for eigenpairs of certain symmetric tridiagonal matrices, \href{http://dx.doi.org/10.1017/S0004972700033918}{Bull. Austral. Math. Soc. \textbf{55}, 249 (1997)}.

\end{thebibliography}
\end{document}